\documentclass{jfm}

\usepackage{graphicx}
\usepackage{natbib}
\usepackage{amsmath}
\usepackage{color}
\usepackage{subfigure}
\usepackage{cancel}

\usepackage{float}
\usepackage{stmaryrd}
\usepackage{mathtools}
\usepackage{float}
\usepackage{array}

\ifCUPmtlplainloaded \else
  \checkfont{eurm10}
  \iffontfound
    \IfFileExists{upmath.sty}
      {\typeout{^^JFound AMS Euler Roman fonts on the system,
                   using the 'upmath' package.^^J}%
       \usepackage{upmath}}
      {\typeout{^^JFound AMS Euler Roman fonts on the system, but you
                   dont seem to have the}%
       \typeout{'upmath' package installed. JFM.cls can take advantage
                 of these fonts,^^Jif you use 'upmath' package.^^J}%
      }
  \else
  \fi
\fi

\ifCUPmtlplainloaded \else
  \checkfont{msam10}
  \iffontfound
    \IfFileExists{amssymb.sty}
      {\typeout{^^JFound AMS Symbol fonts on the system, using the
                'amssymb' package.^^J}%
       \usepackage{amssymb}%
         \let\leq=\leqslant
         \let\geq=\geqslant
      }{}
  \fi
\fi

\ifCUPmtlplainloaded \else
  \IfFileExists{amsbsy.sty}
    {\typeout{^^JFound the 'amsbsy' package on the system, using it.^^J}%
     \usepackage{amsbsy}}
    {\providecommand\boldsymbol[1]{\mbox{\boldmath $##1$}}}
\fi

\newsavebox{\astrutbox}
\sbox{\astrutbox}{\rule[-5pt]{0pt}{20pt}}

\newbox\tempbox

\def\Xint#1{\mathchoice
{\XXint\displaystyle\textstyle{#1}}%
{\XXint\textstyle\scriptstyle{#1}}%
{\XXint\scriptstyle\scriptscriptstyle{#1}}%
{\XXint\scriptscriptstyle\scriptscriptstyle{#1}}%
\!\int}
\def\XXint#1#2#3{{\setbox0=\hbox{$#1{#2#3}{\int}$ }
\vcenter{\hbox{$#2#3$ }}\kern-.6\wd0}}

\def\dashint{\Xint-}

\newcolumntype{L}{>{\centering\arraybackslash}m{3cm}}

\title[Unsteady Aerodynamics of a 2D Airfoil]{Unsteady Aerodynamics and Vortex-sheet Formation of A Two-dimensional Airfoil}

\author[X. Xia and K. Mohseni]%
{  
X.\ns X\ls I\ls A$^{1}$
\ns
\and
K.\ns M\ls O\ls H\ls S\ls E\ls N\ls I$^{1,2}$
  \thanks{Email address for correspondence: mohseni@ufl.edu \hfill \mbox{}}
\ns
}

\affiliation{$^1$Department of Mechanical and Aerospace Engineering.\\[\affilskip]
$^2$Department of Electrical and Computer Engineering.\\[\affilskip]
University of Florida, Gainesville, FL, 32611-6250, USA\\[\affilskip]
}

\begin{document}

\maketitle

\begin{abstract}
Unsteady inviscid flow models of wings and airfoils have been developed to study the aerodynamics of natural and man-made flyers. Vortex methods have been extensively applied to reduce the dimensionality of these aerodynamic models, based on the proper estimation of the strength and distribution of the vortices in the wake. In such modeling approaches, one of the most fundamental questions is how the vortex sheets are generated and released from sharp edges. To determine the formation of the trailing-edge vortex sheet, the classical Kutta condition can be extended to unsteady situations by realizing that a flow cannot turn abruptly around a sharp edge. This condition can be readily applied to a flat plate or an airfoil with cusped trailing edge since the direction of the forming vortex sheet is known to be tangential to the trailing edge. However, for a finite-angle trailing edge, or in the case of flow separation away from a sharp corner, the direction of the forming vortex sheet is ambiguous. To remove any ad-hoc implementation, the unsteady Kutta condition, the conservation of circulation, as well as the conservation laws of mass and momentum are coupled to analytically solve for the angle, strength, and relative velocity of the trailing-edge vortex sheet. The two-dimensional aerodynamic model together with the proposed vortex-sheet formation condition is verified by comparing flow structures and force calculations with experimental results for airfoils in steady and unsteady background flows.
\end{abstract}



   


\section{Introduction}\label{sec:1}

Mankind has been dreaming to fly for centuries. However, the fundamental flying mechanism had not been understood until the pioneers of aerodynamics, such as Kutta and Joukowski \citep{MilneThomsonLM:58a}, connected lift generation to the circulation of an airfoil in the steady sense. Over the last several decades, in order to design high-performance micro aerial vehicles (MAVs), major research effort has been focused on unveiling the unsteady aerodynamic secrets of insects and birds that have demonstrated unrivaled maneuverability and agility. Early researchers \citep{EllingtonCP:84a, Dickinson:93a} have attributed the high lift performance of the natural flyers to an attached leading edge vortex (LEV). Later, numerous experimental investigations \citep{Dickinson:99a,Dickinson:04b,YeoKS:08a,Gharib:10a,DeVoriaAC:12a,DengX:13a,LiuY:15a,WeymouthD:15a,BreuerK:16a} have been carried out to study the dynamics of the wake vortices as well as their effects on force generation for wings or airfoils undergoing unsteady motions, such as accelerating, pitching, flapping, etc. 

For theoretical investigation, inviscid potential flow together with vortex methods has been extensively applied to provide reduced flow model without solving the Navier-Stokes equation. For example, \cite{MinottiFO:02a} adopted a virtual coordinate frame to develop an unsteady framework for a two-dimensional (2D) rotating flat plate and employed a single point vortex to emulate the effect of the LEV. However, the single vortex was still modeled in a quasi-steady manner that the location and circulation of the vortex are fixed during the movement of the plate. \cite{LlewellynSmith:09a}, \cite{EldredgeJD:13a}, and \cite{EldredgeJD:14a} modeled the wake using finite sets of point vortices with varying strengths and evolving locations. This resulted in significant improvement in capturing the unsteady features of the flow; whereas the accuracy of the model is still limited, especially for cases with complex near-field wake patterns, due to the overly-reduced modeling of the vortical structures. An alternative approach is to fully represent the wake vortex sheets in a discretized sense, using either point vortices or vortex panels as demonstrated by \cite{KatzJ:81a}, \cite{TriantafyllouMS:95b}, \cite{JonesMA:03a}, \cite{YuY:03a}, \cite{Pullin:04b}, \cite{AnsariSA:06b, AnsariSA:06a}, \cite{EldredgeJD:07a}, \cite{Mohseni:13y}, \cite{RameshK:14a}, and \cite{WuZ:15a}. Due to a relatively complete representation of all vortical structures in the wake, the vortex-sheet approach generally yields promising accuracy; however, the simulation becomes increasingly expensive as time proceeds. As a remedy, a vortex-amalgamation method \citep{Mohseni:13z,Mohseni:15h} has recently been proposed to effectively restrain the growth of the computational cost for large simulations.      

\begin{figure}
\begin{center}
\scalebox{0.6}{\includegraphics{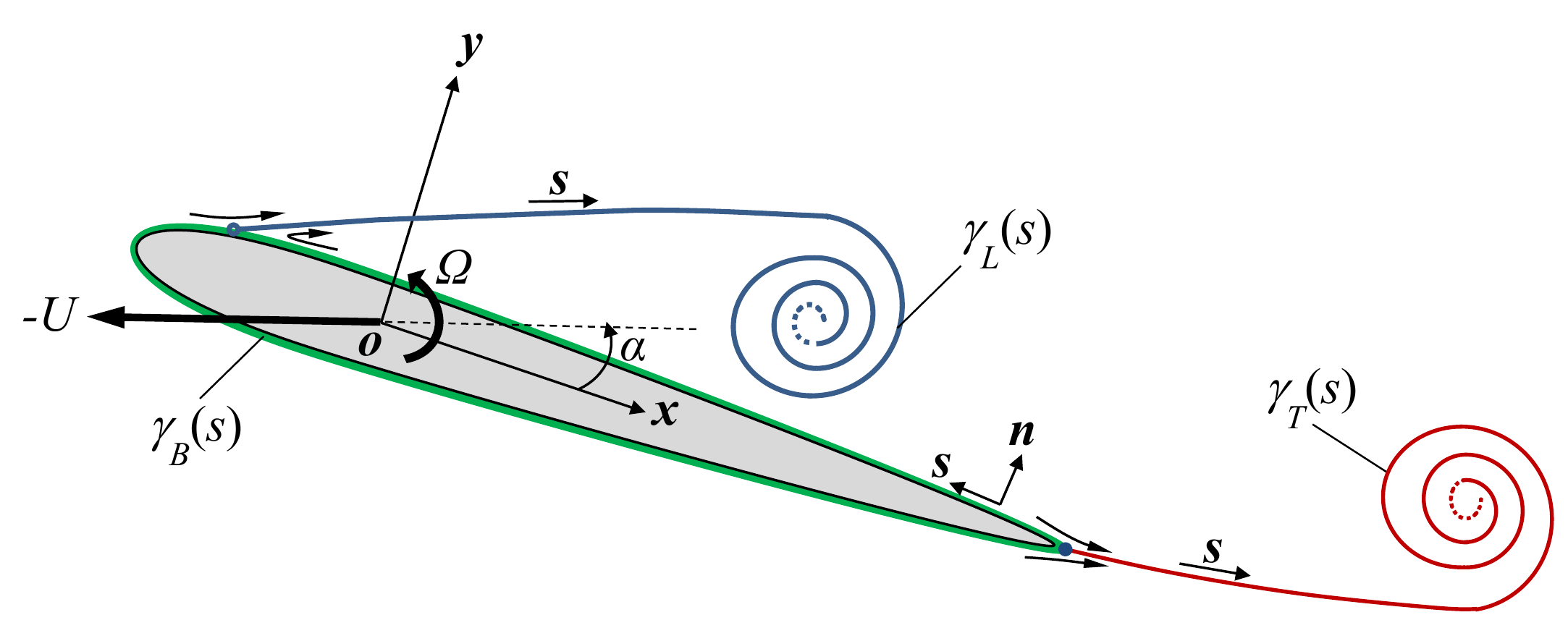}}
\caption{Diagram showing the unsteady flow model of an airfoil.}
\label{fig:physical_flow_AF}
\end{center}
\end{figure}

In practice, our previous model \citep{Mohseni:13y} for a 2D unsteady flat plate could be readily applied to the case of a rigid wing or airfoil with negligible thickness. However, the same extension might not be applicable for an airfoil as the model requires to establish an analytical mapping between the airfoil and a circle. Although special solutions for certain types of airfoil could exist (such as the Joukowski airfoil), it is generally challenging to obtain such transformation for an arbitrary-shaped airfoil. To address this difficulty, the effect of the airfoil might be substituted by a closed vortex sheet coinciding with the surface of the airfoil, the framework of which is consistent with the boundary-element method \citep{MorinoL:74a, KatzJ:81a, KatzJ:91a, TriantafyllouMS:02a, JonesMA:03a, EldredgeJD:07a, YueDKP:12a}. Similar to the flow model for a flat plate, discretized vortex sheets could still be incorporated to account for the wake vortical structures shed from the leading and trailing edges of the airfoil, as illustrated in figure~\ref{fig:physical_flow_AF}.

The essence of vortex-based flow models lies in the accurate predictions of the strength and distribution of the vortices in the flow field. Since the time evolution of free vortical structures can be solved using the Birkhoff-Rott equation \citep{LinCC:41a,BirkhoffG:62a,RottN:56a}, the key problem to be addressed is how vorticity detaches from the surface of the solid body and enters the fluid. In reality, the generation of vorticity is related to the interaction between fluid and solid boundary that forms the shear layer, which is essentially the product of viscous effect. Since the flow model is inviscid, a typical solution to that is applying vorticity releasing conditions at the vortex shedding locations of the solid body, e.g. the Kutta condition at a sharp trailing edge. This means that all the viscous effects can be translated into a single condition \citep{CrightonDG:85a} that yields an estimation of the circulation around the body or the vorticity created near each vortex shedding location. For trailing edges, the classical Kutta condition has been shown to be effective for steady background flows, thus it is also commonly known as the steady-state trailing-edge Kutta condition which requires a finite velocity at the trailing edge \citep{Saffman:77a,ChowCY:81a,MourtosNJ:96a}. For a Joukowski airfoil, the steady state Kutta condition is realized by setting the trailing edge to be a stagnation point in the mapped circle plane. The effect of this implementation is that the stagnation streamline from the trailing edge will be tangential to the edge (or bisect a finite-angle trailing edge), which is consistent with the physical flow near the trailing edge. For the case of a flat plate, this condition will guarantee the streamline emanating from this stagnation point to be inline with the plate, fulfilling the condition proposed in previous studies \citep{ChenSH:87a,PolingDR:87a}. However, the stagnation streamline for a finite-angle trailing edge is ambiguous \citep{PolingDR:86a}, which causes great challenge to modeling the trailing-edge vortex sheet. 

In this study, we employ discretized bound and free vortex sheets to model the unsteady flow around an airfoil. The flow field is given by solving the Euler equation obtained by removing the viscous term in the Navier-Stokes equation. To this end, flow models based on the Euler equation have difficulty in capturing viscous effects around and behind a moving object. The introduction of the vortex sheet could partially address this difficulty. Physically, a vortex sheet represents a viscous shear layer in the Euler limit, by letting the thickness of the shear layer approach zero (section 2.2 of \cite{Saffman:92a}). From a kinematic perspective, this approximation would yield the solution to the inviscid flow outside the vortex sheet with the non-penetration boundary condition implemented at the fluid-solid interface. However, a vortex-sheet is inadequate to represent a viscous shear layer in the dynamic sense. This is because the vortex sheet only conserves the tangential velocity jump, which is also the circulation per unit length of the original shear layer. Therefore, a vortex sheet does not resolve the velocity gradient across the sheet; neither does it account for the mass and momentum associated with the shear layer, nor the fluid entrained by the shear layer. To this end, a vortex-sheet based flow model is likely to capture the force contributions from circulation, i.e. lift and pressure drag, but not the viscous drag which is closely related to the momentum balance of the viscous shear layer. In order to properly capture other viscous effects, such as entrainment, viscous drag, or even energy dissipation, we propose a generalized sheet with superimposed quantities and discontinuities associated with the original shear layer. In this way, the original vortex sheet could be extended to represent a general shear layer at the fluid-fluid or fluid-solid interfaces for single and multiple phase flows.

As seen in this manuscript, application of proper boundary conditions and standard conservation laws to this model allows for the calculation of correct wall-bounded vortex sheet as well as the free vortex sheet released at the trailing edge of an airfoil. The proposed generalized sheet model enables the application of the conservation laws of mass and momentum for a system of triple-joint vortex sheets and surrounding flow. The result will be applied to a particular case, which is the finite-angle trailing edge of an airfoil where two incoming bound vortex sheets on the airfoil surface join together to form the free vortex sheet. Together with the unsteady Kutta condition and the conservation of circulation, one obtains a general analytical condition to determine the angle, strength, and relative velocity of the trailing-edge vortex sheet. 

\section{Unsteady Flow Model}\label{sec:2}

The framework of the flow model for a two-dimensional (2D) airfoil is not fundamentally different from that for a 2D flat plate wing \citep{Mohseni:13y}. In both situations, potential flow is applied as the governing equation, which is based on solving the Navier-Stokes equation in the Eulerian limit. This has two main advantages: one is analytical representation of the entire flow field, the other is saving computational cost since the domain of interest is reduced from the entire flow field to only finite vortical structures. 

Assuming that the rigid-body motion of the airfoil in a quiescent environment can be decomposed into a translational motion of velocity $-U(t)$ and a rotational motion of angular velocity $\Omega(t)$. Both the translational and the rotational motions can be incorporated into the boundary condition at the solid-fluid interface. As shown in figure~\ref{fig:physical_flow_AF}, flow separation near the leading edge and at the sharp trailing edge of the airfoil causes the formation of two free vortex sheets in the wake. In a Cartesian coordinate system with the origin fixed at the rotation center, the complex potential of the flow around an airfoil with angle of attack, $\alpha(t)$, can be formulated as
\begin{equation}
\label{eq: cplxpotential_AF0}
\begin{split}
 w(z,t) & = - \frac{i}{2\pi} \left[ \vphantom{\int} \right. \underbrace{\int_{0}^{S_{L}(t)} \ln \left( z - z_{L}(s,t) \right) \gamma_{L}(s,t) \mathrm{d} s}_{\text{LEV term}}\\ 
& + \underbrace{\int_{0}^{S_{T}(t)} \ln \left( z - z_{T}(s,t) \right) \gamma_{T}(s,t) \mathrm{d} s}_{\text{TEV term}} \left. \vphantom{\int} \right]  + \underbrace{\vphantom{\int} w_b(z,t)}_{\text{Body term}},
\end{split}
\end{equation}
where $z$ is the complex position, $s$ is the curve length between the separation point and a vortex element along a vortex sheet, $S$ represents the total length of an entire vortex sheet, and $\gamma$ is the vortex sheet strength (circulation per unit length). The subscripts $_{L}$ and $_{T}$ denote the properties associated with the leading-edge and trailing-edge vortex sheets, respectively. Here, $w_b(z,t)$ represents the flow induced by the body motion of the airfoil, and is usually associated with the so-called `bound vortex'. Therefore, the `bound vortex' can be viewed as a substitute for the solid body so that the non-penetration boundary condition can still be satisfied at the fluid-solid interface while the solid body is removed from the flow model. Again, we note here that the `body term' or the `bound vortex' implicitly accounts for the effects of translation, rotation, or deformation, and more details will be provided in Section~\ref{sec:3}. In general, `bound vortex' can be realized by placing image vortices inside the solid body for a Joukowski airfoil or a flat plate, where the strength and location of the image vortices can be first decided from Milne-Thomson's circle theorem \citep{MilneThomsonLM:58a} in the circle plane and then mapped back to the physical plane. However, for an arbitrarily-shaped airfoil which can not be easily mapped to a circle, an analytical solution for $w_b(z,t)$ is not available. In this case, the `bound vortex' can be realized by placing a bound vortex sheet along the surface of the airfoil as shown in figure~\ref{fig:physical_flow_AF}, and $w_b(z,t)$ becomes
\begin{equation}
\label{eq: cplxpotential_AF1}
 w_b(z,t) = -\frac{i}{2\pi} \int_{0}^{S_{B}(t)} \ln \left( z - z_{B}(s,t) \right) \gamma_{B}(s,t) \mathrm{d} s,
\end{equation}
where the subscript $_{B}$ denotes the properties associated with the bound vortex sheet. Note here that $s$ for the bound vortex sheet starts from the trailing edge with a counter-clockwise direction. Now, combining equations~(\ref{eq: cplxpotential_AF0}) and~(\ref{eq: cplxpotential_AF1}) and taking the derivative $\mathrm{d} w/\mathrm{d} z$, we obtain the complex-conjugate velocity field, $\bar{V}(z,t) = u(z,t) - iv(z,t)$, in the form 
\begin{equation}
\label{eq: cplxvelocity_AF0}
\begin{split}
 \bar{V}(z,t) & = - \frac{i}{2\pi} \left[ \vphantom{\int} \right. \underbrace{\int_{0}^{S_{L}(t)} \frac{\gamma_{L}(s,t) \mathrm{d} s}{z - z_{L}(s,t)}}_{\text{LEV term}} + \underbrace{\int_{0}^{S_{T}(t)} \frac{\gamma_{T}(s,t) \mathrm{d} s}{z - z_{T}(s,t)}}_{\text{TEV term}}\\
 & + \underbrace{\int_{0}^{S_{B}(t)} \frac{\gamma_{B}(s,t) \mathrm{d} s}{z - z_{B}(s,t)}}_{\text{Bound vortex sheet term}} \left. \vphantom{\int} \right].
\end{split}
\end{equation}
It should be noted that the velocity field represented by equation~(\ref{eq: cplxvelocity_AF0}) is singular on the vortex sheets, where the jump of the tangential-component velocity is equal to the strength of the vortex sheet \citep{Saffman:92a}. More details regarding the evaluation of the vortex sheets will be discussed in sections~\ref{sec:3} and~\ref{sec:4}. At this point, the calculation of the entire flow field are reduced to determining the strength and distribution of only a few finite-length vortex sheets.   

Following previous studies \citep{WuJC:81a,EldredgeJD:10a}, the aerodynamic force applied on the airfoil can be estimated based on the rate of change of the total impulse in the form
\begin{equation}
\label{eq:AF_force}
\boldsymbol{F} = - \rho \frac{\mathrm{d}}{\mathrm{d}t} \int_{\sum S} \boldsymbol{x} \times \boldsymbol{\gamma}\mathrm{d}s,
\end{equation}
where $\boldsymbol{x}$ is the position vector of a vortex-sheet element, and $\boldsymbol{\gamma} = \gamma \hat{\boldsymbol{k}}$, where $\hat{\boldsymbol{k}}$ is the unit vector normal to the 2D plane. $\rho$ is the density. $\sum S$ represents the entire vortex-sheet system, and $\sum S = S_{L} + S_{T} + S_{B}$ in the current model. Similarly, the total torque exerted by the fluid on the airfoil can be obtained from
\begin{equation}
\label{eq:AF_torque}
T_{\tau} = - \rho \frac{\mathrm{d}}{2\mathrm{d}t} \int_{\sum S} \boldsymbol{x} \times (\boldsymbol{x} \times \boldsymbol{\gamma}\mathrm{d}s).
\end{equation}  
The main advantage of equations~(\ref{eq:AF_force}) and~(\ref{eq:AF_torque}) is that the calculations of force and torque are completely transformed into the dynamics of the bound and wake vorticies, which can be explicitly obtained from this aerodynamic model.
\section{Bound Vortex Sheet}\label{sec:3}

The instantaneous velocity field around an airfoil can now be decided if the two free vortex sheets and one bound vortex sheet are given. This requires knowing the strengths and positions of the vortex sheets ($\gamma_{L}, \gamma_{T}, \gamma_{B}, z_{L}, z_{T}, z_{B}$). Considering the case where the flow initially remains fully attached, this indicates no flow separation or free vortex sheet existed at $t=0$. Under this assumption, $\gamma_{L}$, $\gamma_{T}$, $z_{L}$, $z_{T}$ for later times might be found through solving the formation and evolution of the free vortex sheets. So we assume that $\gamma_{L}$, $\gamma_{T}$, $z_{L}$, $z_{T}$ are known in order to solve the bound vortex sheet at any given time. Furthermore, the position of the bound vortex sheet, $z_{B}$, is also known as it coincides with the surface of the airfoil at any time. As a result, the main task here is to solve for the vortex sheet strength $\gamma_{B}$. We should note that a bound vortex sheet is treated differently from a free vortex sheet since $z_{B}$ is prescribed. Actually, the free vortex sheet is applied to represent the physical free shear layer, while the bound vortex sheet is introduced to `mimic' the effect of solid boundary. Therefore, it is expected that the primary role of the bound vortex sheet is to satisfy the non-penetration boundary condition, which can be expressed as  
\begin{equation}
\label{eq: BC_AF}
 \boldsymbol{u}(z') \cdot \hat{\boldsymbol{n}}(z') = \boldsymbol{u}_b(z') \cdot \hat{\boldsymbol{n}}(z') \hspace{5mm}\text{for}\hspace{5mm} z' = z_{B}(s') \hspace{2mm}\text{and}\hspace{2mm} 0 \leq s' \leq S_{B},
\end{equation}
where $\boldsymbol{u}(z') = (u(z'),v(z'))$ is the flow velocity at the surface of the airfoil, $z_{B}$, and $\hat{\boldsymbol{n}}(z')$ is the unit normal vector of the surface. Note that the definitions for $z'$ and $s'$ only applies to Sections~\ref{sec:3} and~\ref{sec:4}. Also, time $t$ is dropped here and in following derivations for simplicity although they should be satisfied instantaneously. $\boldsymbol{u}_b(z')$ is the velocity associated with the surface element of the airfoil so it generally describes the deformation of an airfoil. However, $\boldsymbol{u}_b(z')$ can be also applied to account for the translational motion in the complex-conjugate form, $-|U|e^{-i\alpha}$, and the rotational motion in the complex-conjugate form, $-i \Omega \bar{z}'$, where $\bar{z}'$ denotes the complex conjugate of $z'$. Since the bound vortex sheet is placed at the surface of the airfoil, it creates a velocity jump across $z_{B}$. Based on equation~(\ref{eq: cplxvelocity_AF0}) and the definition of a vortex sheet \citep{Saffman:92a}, the two limiting values for $u^{\pm}(z') - iv^{\pm}(z') = \bar{V}_B^{\pm}(z')$ can be derived as 
\begin{equation}
\label{eq: BC_vel0}
\begin{split}
 \bar{V}_B^{\pm}(z') & = - \frac{i}{2\pi} \left[ \int_{0}^{S_{L}} \frac{\gamma_{L}(s) \mathrm{d} s}{z' - z_{L}(s)} + \int_{0}^{S_{T}} \frac{\gamma_{T}(s) \mathrm{d} s}{z' - z_{T}(s)} \right.\\ 
& + \left. \dashint_{0}^{S_{B}} \frac{\gamma_{B}(s) \mathrm{d} s}{z' - z_{B}(s)} \right] \pm \frac{1}{2}\gamma_{B}(s')\frac{\mathrm{d} \bar{z}'}{|\mathrm{d} z'|},
\end{split}
\end{equation}
where $\dashint$ denotes the Cauchy principal value which excludes the vorticity at $z'$ from the integral. $\mathrm{d} z' |\mathrm{d} z'|^{-1}$ is the complex form of the unit tangential vector, $\hat{\boldsymbol{s}}(z')$, at the surface of the airfoil. With $\hat{\boldsymbol{s}}(z')$ pointing in the counter-clockwise direction of the airfoil body, $\bar{V}_B^{+}(z')$ becomes the velocity limit when the bound vortex sheet is approached from the outside of the airfoil, whereas $\bar{V}_B^{-}(z')$ is the velocity limit when the vortex sheet is approached from the inside. Since $\boldsymbol{u}(z')$ is the flow velocity outside the surface of the airfoil, it should take the value $\bar{V}_B^{+}(z')$. With $\hat{\boldsymbol{n}}(z')$ written as $-i \mathrm{d} z' |\mathrm{d} z'|^{-1}$, equation~(\ref{eq: BC_AF}) has the complex form
\begin{equation}
\label{eq: BC_AF1}
 \text{Re} \left\{ \left[\bar{V}_B^{+}(z') + |U|e^{-i\alpha} + i \Omega \bar{z}' \right] \frac{-i \mathrm{d} z'}{|\mathrm{d} z'|} \right\} = 0.
\end{equation}
Ideally, equation~(\ref{eq: BC_AF1}) would give the strength of the bound vortex sheet, $\gamma_{B}$, if $\gamma_{L}$, $\gamma_{T}$, $z_{L}$, $z_{T}$, and $z_{B}$ are given. However, a general analytical solution to equation~(\ref{eq: BC_AF1}) does not exist for an arbitrarily-shaped airfoil. Fortunately, it is possible to solve this problem numerically by discretizing the bound vortex sheet into piecewise linear vortex panels. 

It should be noted that the strength of the bound vortex sheet $\gamma_{B}$ can be expressed as $\gamma_{B} = \boldsymbol{u}_f \cdot \hat{\boldsymbol{s}}$, where $\boldsymbol{u}_f$ represents the potential flow velocity at the fluid-solid boundary. With no-slip boundary condition, $\gamma_{B}$ can be divided into two terms, $\gamma_{b}$ and $\gamma_{\gamma}$, according to \cite{EldredgeJD:10a}. $\gamma_{b}$ is purely associated with the body-surface motion relative to the reference frame, and it can be estimated from $\gamma_{b} = \boldsymbol{u}_b \cdot \hat{\boldsymbol{s}}$. $\gamma_{\gamma}$ is the physical vortex sheet corresponding to the viscous shear layer, which is given by $\gamma_{\gamma} = \gamma_{B} - \gamma_{b}$. Therefore, $\gamma_{\gamma}$ is invariant regardless of the reference frame being global or body-fixed, while both $\gamma_{b}$ and $\gamma_{B}$ could change as the reference frame changes. To avoid ambiguity, $\gamma_{B}$ in this study only represents the bound vortex sheet in the global reference frame.  

\section{Formation of Free Vortex Sheets}\label{sec:4}

Now, with the velocity field and the bound vortex sheet determined from equations~(\ref{eq: cplxvelocity_AF0}) and~(\ref{eq: BC_AF1}), respectively, we are faced with the task of determining the intensities and locations of the two free vortex sheets since $\gamma_{L}$, $\gamma_{T}$, $z_{L}$, and $z_{T}$ are the prerequisites for both equations. Note again that the flow is assumed to be fully attached at $t = 0$, which means there is no free vortex sheet initially. Therefore, determining the vortex sheets in the wake at $t > 0$ requires understanding of the formation and evolution of the free vortex sheets that are detached from the airfoil. 

The evolution of a free vortex sheet should follow Helmholtz laws of vortex motion \citep{HelmholtzH:1867a, Saffman:92a} for barotropic fluid with conservative body force. According to the third Helmholtz law, the circulation of a vortex sheet element can be treated as time invariant once it is detached from the airfoil. Furthermore, the second Helmholtz law dictates that a vortex element and its overlapping fluid particle should move together in the flow. In accordance with these principles, the velocity describing the motion of an element on a free vortex sheet can be derived using the Birkhoff-Rott equation \citep{LinCC:41a, BirkhoffG:62a, RottN:56a}. As a result, this velocity formulation is similar to equation~(\ref{eq: cplxvelocity_AF0}), with $\int_{0}^{S_{L}(t)}$ replaced by $\dashint_{0}^{S_{L}(t)}$ for an element on the leading-edge vortex sheet or $\int_{0}^{S_{T}(t)}$ replaced by $\dashint_{0}^{S_{T}(t)}$ for an elment on the trailing-edge vortex sheet. Again, $\dashint$ denotes the Cauchy principle integral, which removes the singularity induced by a vortex element itself. Now, with the instantaneous distributions of vorticity obtained from the vortex-sheet evolution, we are left with the question of how vorticity is generated and shed from the root of a free vortex sheet. 

\subsection{The challenge with a finite-angle trailing edge}\label{sec:4-1}

We first consider a simple case, where the vortex sheet is formed at the edge of a flat plate or a cusped trailing edge of an airfoil. Without considering the viscous effect, a typical way of deciding the vortex-sheet formation at the trailing edge is the classical steady Kutta condition. This condition requires the flow velocity at the trailing edge to be finite or the loading at the trailing edge to be zero, based on the physical sense that flow cannot turn around a sharp edge. The application of this condition for a flat plate or a Joukowski airfoil (with cusped trailing edge) has already been demonstrated in several previous works \citep{TriantafyllouMS:95b, YuY:03a, AnsariSA:06a, Mohseni:13y} among others. Basically, this condition is equivalent to enforcing a stagnation point at the trailing edge in the transformed circle plane. However, \cite{Mohseni:14i} recently pointed out that a stagnation point generally does not exist at the trailing edge for the case of body rotation. As a result, they proposed to implement the unsteady Kutta condition by relaxing the trailing edge point of the circle plane from totally stagnant to only stagnant in the tangential direction of the surface, which still conforms to the requirement of the classical Kutta condition in the sense of preventing flow around the sharp edge. Again, it is emphasized here that these steady and unsteady Kutta conditions should be implemented in the transformed circle plane, which means they only apply to an airfoil that can be mathematically mapped to a circle.             

Alternatively, \cite{JonesMA:03a} modeled the flow around a flat plate using a bound vortex sheet coincident with the plate and two free vortex sheets that are emanating from the plate's two sharp edges, which is similar to the flow model presented here for an airfoil. By removing the singularities of the flow velocity at the trailing edge, which complies with the classical Kutta condition that flow velocity should be finite at a sharp edge, Jones managed to derive an analytical formulation for the unsteady Kutta condition as
\begin{equation}
\label{eq: FP_KC}
\dot{\Gamma}_{g} = \frac{\partial \Gamma_{g}}{\partial t} = u_{E}\gamma_{E},
\end{equation}     
where $\Gamma_{g}$ is the total circulation of the forming vortex sheet, so $\dot{\Gamma}_{g}$ is the rate at which circulation is generated at the sharp edge to form the free vortex sheet. $u_{E}$ represents the average tangential slip between the plate and the bound vortex sheet at the sharp edge, and $\gamma_{E}$ is the strength of the bound vortex sheet at the edge. It is important to note that $u_{E}$ and $\gamma_{E}$ are properties associated with the bound vortex sheet. For an unsteady flow, according to the study of \cite{WuJZ:06a} (eq. 4.134), the free vortex sheet formed at the sharp edge satisfies $\partial \Gamma_{g}/\partial t = u_{g}\gamma_{g}$, where $u_{g}$ and $\gamma_{g}$ are the tangential velocity component and the strength of the forming vortex sheet, respectively. A Comparison between this equation and equation~(\ref{eq: FP_KC}) suggests that the strength of the forming vortex sheet is equal to the strength of its adjacent bound vortex sheet, while the tangential velocity of the forming vortex sheet relative to the flat plate equals the average tangential slip velocity between the bound vortex sheet and the sharp edge. Furthermore, Jones' derivation also suggests that the tangential directions of the forming vortex sheet and the bound vortex sheet should match at the flow separation edge in order to completely remove the velocity singularities. Therefore, Jones' unsteady Kutta condition allows the analytical calculation of the direction, velocity, and strength of the forming vortex sheet for the trailing edge of a flat plate or a cusped airfoil, without any arbitrary implementation in the shedding procedure. 

\begin{figure}
 \center{\includegraphics[width=0.85\textwidth]{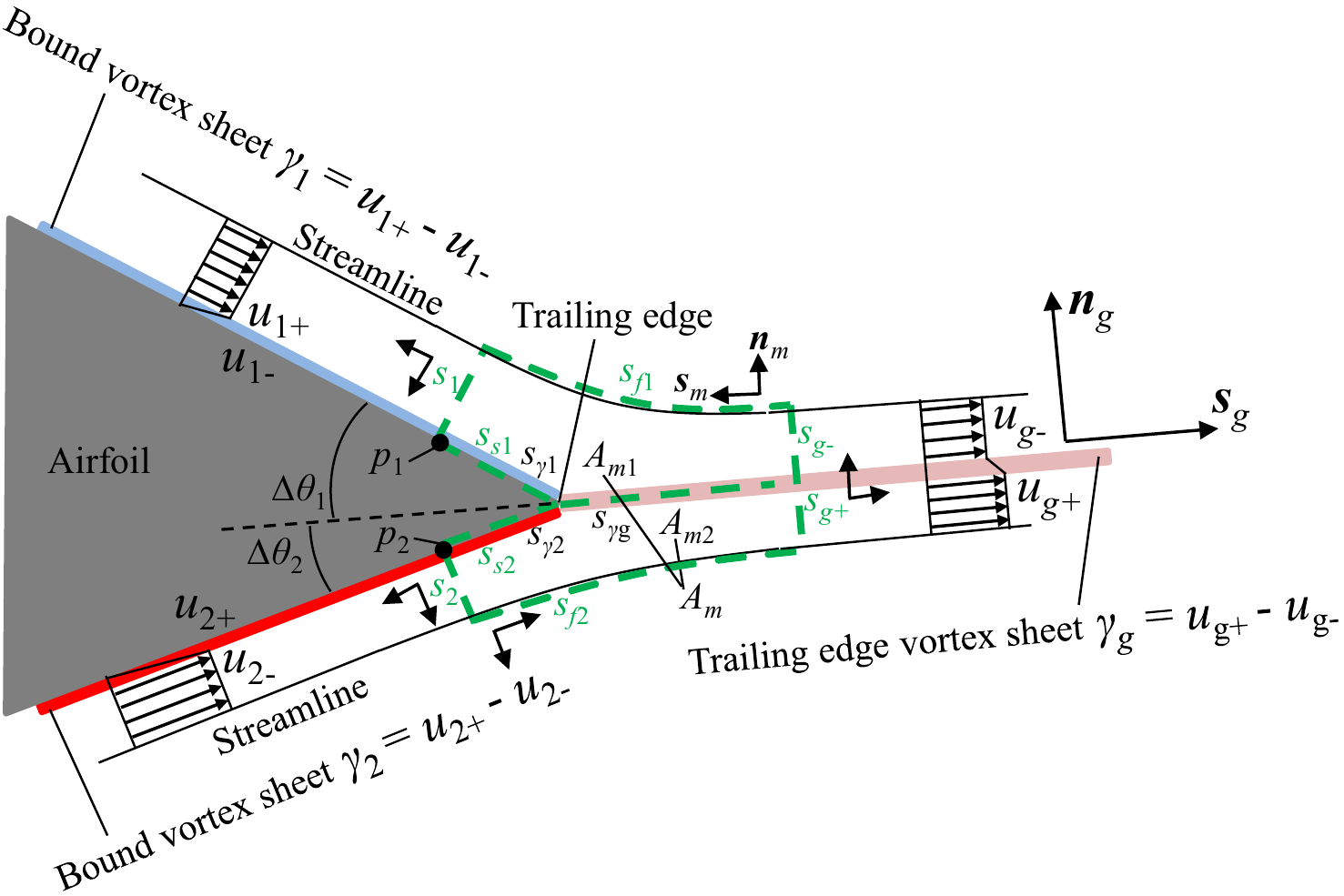}}
 \caption{\small The formation of a free vortex sheet at a finite-angle trailing edge. The green dashed lines ($S_{s1}$, $S_{1}$, $S_{f1}$, $S_{g-}$, $S_{g+}$, $S_{f2}$, $S_{2}$, and $S_{s2}$) together form the boundary of a material volume $A_m$, where the flow on both sides of the trailing edge merge into one stream and form a free vortex sheet. $A_m$ can be divided into two sub-volumes, $A_{m1}$ and $A_{m2}$, by the forming vortex sheet. The velocities associated the vortex sheets ($u_{1-}$, $u_{1+}$, $u_{2-}$, $u_{2+}$, $u_{g-}$, and $u_{g+}$) are normal velocities defined based on the surfaces of $A_m$ as $u_n = \boldsymbol{u} \cdot \hat{\boldsymbol{n}}_m$.}
 \label{fig:TE-schm}
\end{figure}

Since the current work is concerned with a general-shaped airfoil, the trailing edge of which could be different from a flat plate or a cusped edge like the Joukowski airfoil, Jones' unsteady Kutta condition might not be suitable. Specifically, if the upper and lower surfaces of the trailing edge have different tangential directions as shown in figure~\ref{fig:TE-schm}, the vortex-sheet configuration would be fundamentally different from that of Jones' work. To generalize this problem, we consider a sharp trailing edge where there is a finite angle, $\Delta \theta_0 \in [0,\pi)$, between the upper and lower surfaces. A relevant question here is how to decide the direction of the forming free vortex sheet, given that the direction of the bound vortex sheet is ambiguous at the trailing edge. 

Before further discussions, here we emphasize that the second Helmholtz law of vortex motion \citep{HelmholtzH:1867a, Saffman:92a} dictates that a free vortex sheet moves with its background flow as a material sheet. From a kinematic perspective, this means that a forming vortex sheet could be viewed as a streakline released from the vortex shedding edge in the body-fixed reference frame. Furthermore, at the releasing location of a streakline, the directions of the streakline and the streamline are identical to each other. This further indicates that the direction of the forming vortex sheet coincides with the direction of the stagnation streamline at the trailing edge in the body-fixed reference frame. Therefore, the ambiguity of the vortex-sheet direction is also reflected by the ambiguity of the streamline direction, which has been investigated by many previous studies. Actually, for steady trailing-edge flow where the shedding of vorticity vanishes ($\dot{\Gamma}_g = 0$), \cite{PolingDR:86a} pointed out that the steady Kutta condition requires the stagnation streamline to bisect the wedge angle of a finite-angle trailing edge. Otherwise, an unbalance between the upper and lower shear layers near the trailing edge would cause a non-zero vorticity generation which would violate the steady flow condition. According to this argument, an unsteady trailing-edge flow naturally generates vorticity and causes the stagnation streamline to divert from the wedge bisector line, which has been confirmed experimentally \citep{HoCM:81a, PolingDR:86a}. A prominent theory for the unsteady situation has been proposed by \cite{GiesingJP:69a} and \cite{MaskellEC:71a} that the stagnation streamline is an extension of one of the two tangents at the trailing edge. Although \cite{BasuBC:77a} has provided extensive discussion supporting the Giesing-Maskell model, a notable drawback of this model is that it does not reduce to the steady-state solution in the limit of $\dot{\Gamma}_g \rightarrow 0$. Furthermore, \cite{PolingDR:86a} reported that the Giesing-Maskell model only approximately holds for large $\dot{\Gamma}_g$, while they observed a smooth change of the stagnation-streamline direction for small $\dot{\Gamma}_g$.            

In this study, we believe that the generation of the free vortex sheet at the trailing edge not only satisfies the physical kinematic condition, i.e. the Kutta condition, but also complies with the conservation laws of circulation, mass, and momentum associated with the shear layers and their surrounding flow. Figure~\ref{fig:TE-schm} shows the merging process of the upper and lower bound vortex sheets, which result in the formation of a free vortex sheet at the trailing edge. To formulate the problem, we define a 2D material volume $A_m$ in the body-fixed reference frame with its boundary $\partial A_m = S_{s1} + S_{1} + S_{f1} + S_{g-} + S_{g+} + S_{f2} + S_{2} + S_{s2}$. $\hat{\boldsymbol{s}}_m$ and $\hat{\boldsymbol{n}}_m$ are the unit tangential and normal vectors of $\partial A_m$, respectively. And we recall that $\hat{\boldsymbol{s}}$ and $\hat{\boldsymbol{n}}$ are the unit tangential and normal vectors of a vortex sheet. $\boldsymbol{u}$ and $\boldsymbol{\omega}=\omega \hat{\boldsymbol{k}}$ represent the velocity and the vorticity, respectively. A few physical assumptions and boundary conditions are listed below to simplify this problem.\\
(a) The merging zone $A_m$ in reality should be a finite volume (area) with a length scale of $\epsilon_s$. So $S_{1},S_{f1},S_{g-},S_{g+},S_{f2},S_{2}$ have the dimension of $\mathrm{O}(\epsilon_s)$. The merging process does not happen until the upper and lower streams meet exactly at the trailing edge, and any lead area of $A_m$ before the trailing edge should be much smaller than $A_m$ itself. To this point, the length scale of $S_{s1}$ and $S_{s2}$ are assumed to be $\mathrm{o}(\epsilon_s)$. For approximated solution, the length scale $\epsilon_s$ will be assumed to approach zero in the final derivations of this study.\\ 
(b) $S_{f1}$ and $S_{f2}$ coincide with streamlines, so there is no mass flux across the surfaces and $u_n = \boldsymbol{u} \cdot \hat{\boldsymbol{n}}_m = 0$.\\
(c) Assuming the flow field changes smoothly, so $\partial/\partial t$ of any quantity is finite.

In the following we first take steps to derive the Kutta condition and conservation of circulation for the above described problem. The conservations of mass and momentum will be discussed in Section~\ref{sec:5}.

\subsection{Unsteady Kutta condition}\label{sec:4-2}

We start with the analytical implementation of the physical kinematic relation, the unsteady Kutta condition. According to our previous study of an unsteady flat plate \citep{Mohseni:13y, Mohseni:14i}, the rate of circulation generation $\dot{\Gamma}_{g}$ at the trailing edge can be calculated by satisfying the condition
\begin{equation}
\label{eq: unsteady_Kutta}
  \boldsymbol{u}_{g} \cdot \hat{\boldsymbol{n}}_{g} = 0,
\end{equation}
which enforces the streamline in the tangential direction of the forming vortex sheet. For a flat plate or a cusped airfoil, this condition basically requires the vortex sheet to be tangential to the trailing-edge surface, which is consistent with the classical Kutta condition that flow cannot turn around a sharp edge. As discussed in Section~\ref{sec:4-1}, the rationale for equation~(\ref{eq: unsteady_Kutta}) is that the vortex sheet can be viewed as a streakline in the body-fixed reference frame, which coincides with the streamline at the trailing edge. Here, equation~(\ref{eq: unsteady_Kutta}) will extended to the situation of a finite-angle trailing edge, with the direction of the forming vortex sheet assumed to be known for the time being.
\begin{figure}
 \center{\includegraphics[width=0.99\textwidth]{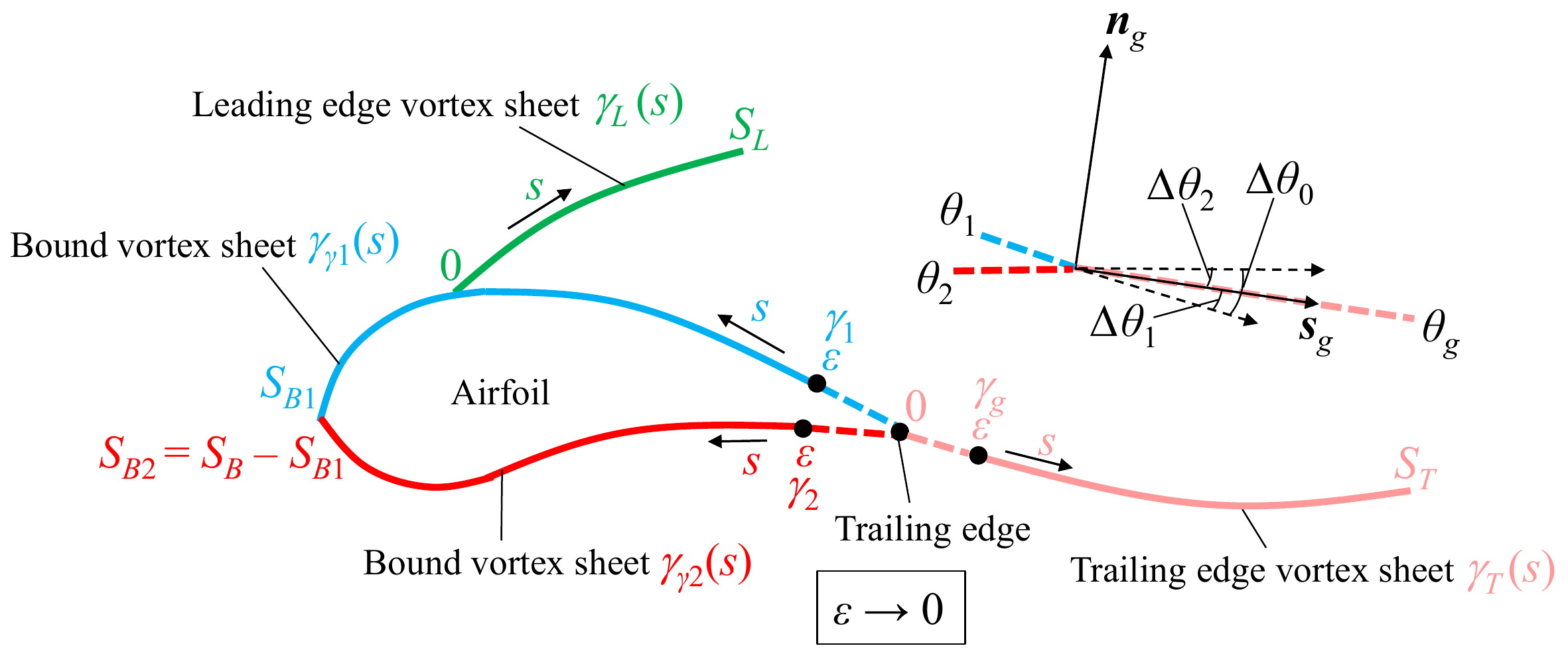}}
 \caption{\small The vortex-sheet configuration for equation~(\ref{eq: VS_vel_TE1}).}
 \label{fig:TE-VS-schm}
\end{figure}

To implement equation~(\ref{eq: unsteady_Kutta}), one needs first to calculate $\boldsymbol{u}_{g}$, the flow velocity at the trailing edge. Figure~\ref{fig:TE-VS-schm} illustrates the vortex-sheet structures near the trailing edge, where $\gamma_1$ and $\gamma_2$ are the bound vortex strengths and $\gamma_g$ is the strength of the forming vortex sheet as they approaches the trailing edge. We start by noting that the vortex-sheet strength is not well defined at the trailing-edge point, where $\gamma_1$, $\gamma_2$, and $\gamma_g$ are discontinuous with each other. So the trailing-edge point is actually a singularity point in the vortex-sheet system. Fortunately, according to the Birkhoff-Rott equation \citep{LinCC:41a, BirkhoffG:62a, RottN:56a}, $\boldsymbol{u}_{g}$ is estimated based on the de-singularized flow field without considering the vortex at the trailing edge point. Based on the vortex-sheet configuration of figure~\ref{fig:TE-VS-schm} and equation~(\ref{eq: cplxvelocity_AF0}), $\boldsymbol{u}_{g}$ can be expressed in the limit form
\begin{equation}
\label{eq: VS_vel_TE1}
\begin{split}
 \bar{V}_{g} & = - \frac{i}{2\pi} \lim_{\epsilon\rightarrow0} \left[ \int_{0}^{S_{L}} \frac{\gamma_{L}(s) \mathrm{d} s}{z_{T}(0) - z_{L}(s)} + \int_{\epsilon}^{S_{T}} \frac{\gamma_{T}(s) \mathrm{d} s}{z_{T}(0) - z_{T}(s)} \right.\\
& + \left. \int_{\epsilon}^{S_{B1}} \frac{\gamma_{\gamma1}(s) \mathrm{d} s}{z_{T}(0) - z_{B1}(s)} + \int_{\epsilon}^{S_{B2}} \frac{\gamma_{\gamma2}(s) \mathrm{d} s}{z_{T}(0) - z_{B2}(s)} \right] + \bar{V}_{CT},
\end{split}
\end{equation}
where $\bar{V}_{CT}$ is the velocity difference associated with the coordinate transformation from the global reference frame to the body-fixed reference frame. $t$ in equation~(\ref{eq: cplxvelocity_AF0}) is dropped here for brevity. Recall the discussion of the bound vortex sheet in Section~\ref{sec:3}, $\gamma_{\gamma}(s)$ rather than $\gamma_{B}(s)$ should be used here for velocity calculation because $\gamma_{b}(s)=0$ in the body-fixed reference frame. For simplicity, we further divide $\gamma_{\gamma}(s)$ into two parts, $\gamma_{\gamma1}(s)$ and $\gamma_{\gamma2}(s)$, as shown in figure~\ref{fig:TE-VS-schm}. The relationships between the original and the divided bound vortex sheets are given by $\gamma_{\gamma1}(s) = \gamma_{\gamma}(s)$ and $z_{B1}(s) = z_{B}(s)$ for $0<s\leq S_{B1}$, and $\gamma_{\gamma2}(s) = \gamma_{\gamma}(S_{B}-s)$ and $z_{B2}(s) = z_{B}(S_{B}-s)$ for $0<s\leq (S_{B}-S_{B1})$, where $S_{B1}$ and $S_{B2}$ satisfy $S_{B1}+S_{B2}=S_{B}$. In this way, the two bound vortex sheets both `stem' from the trailing edge, meaning $\lim_{\epsilon\rightarrow0} z_{B1}(\epsilon) = \lim_{\epsilon\rightarrow0} z_{B2}(\epsilon)$, and $\lim_{\epsilon\rightarrow0} \gamma_{\gamma1}(\epsilon) = \gamma_1$ and $\lim_{\epsilon\rightarrow0} \gamma_{\gamma2}(\epsilon) = \gamma_2$.

To evaluate equation~(\ref{eq: VS_vel_TE1}), the main challenge is that the integrands of $\int_{\epsilon}^{S_{T}}$, $\int_{\epsilon}^{S_{B1}}$, and $\int_{\epsilon}^{S_{B2}}$ become singular as $\epsilon\rightarrow0$. The solution to this is provided in appendices~\ref{sec:10} and~\ref{sec:11}, based on the assumption: finite values, $\epsilon_1$, $\epsilon_2$, and $\epsilon_{T}$, exist so that $\gamma_{\gamma1}(s)$ and $z_{B1}(s)$ are smooth for $0<s\leq \epsilon_1$, $\gamma_{\gamma2}(s)$ and $z_{B2}(s)$ are smooth for $0<s\leq \epsilon_2$, and $\gamma_{T}(s)$ and $z_{T}(s)$ are smooth for $0<s\leq \epsilon_2$. For $z_{B1}(s)$ and $z_{B2}(s)$, their smoothness is related to the shape of the airfoil and can be readily justified for a finite-angle trailing edge. For $\gamma_{\gamma1}(s)$, $\gamma_{\gamma2}(s)$, $\gamma_{T}(s)$, and $z_{T}(s)$, their smoothness should be guaranteed if the entire flow field changes smoothly. With this setup, equation~(\ref{eq: VS_vel_TE1}) can be written as
\begin{equation}
\label{eq: VS_vel_TE2}
\begin{split}
 \bar{V}_{g} & = - \frac{i}{2\pi} \lim_{\epsilon\rightarrow0} \left[ \int_{\epsilon}^{\epsilon_T} \frac{\gamma_{T}(s) \mathrm{d} s}{z_{T}(0) - z_{T}(s)} + \int_{\epsilon}^{\epsilon_1} \frac{\gamma_{\gamma1}(s) \mathrm{d} s}{z_{T}(0) - z_{B1}(s)} \right. \\ 
& \left. + \int_{\epsilon}^{\epsilon_2} \frac{\gamma_{\gamma2}(s) \mathrm{d} s}{z_{T}(0) - z_{B2}(s)} \right] - \frac{i}{2\pi} \left[ \int_{0}^{S_{L}} \frac{\gamma_{L}(s) \mathrm{d} s}{z_{T}(0) - z_{L}(s)} + \int_{\epsilon_T}^{S_{T}} \frac{\gamma_{T}(s) \mathrm{d} s}{z_{T}(0) - z_{T}(s)} \right. \\ 
& \left. + \int_{\epsilon_1}^{S_{B1}} \frac{\gamma_{\gamma1}(s) \mathrm{d} s}{z_{T}(0) - z_{B1}(s)} + \int_{\epsilon_2}^{S_{B2}} \frac{\gamma_{\gamma2}(s) \mathrm{d} s}{z_{T}(0) - z_{B2}(s)} \right] + \bar{V}_{CT}.
\end{split}
\end{equation}
Applying appendix~\ref{sec:10} to the first three integrals and appendix~\ref{sec:11} to the last four integrals yields
\begin{equation}
\label{eq: VS_vel_TE3}
 \bar{V}_{g} = - \frac{i}{2\pi} \lim_{\epsilon\rightarrow0} \left[ \gamma_g e^{-i\theta_g} \ln \left(\epsilon\right) + \gamma_1 e^{-i\theta_1} \ln \left(\epsilon\right) + \gamma_2 e^{-i\theta_2} \ln \left(\epsilon\right) \right] + \bar{V}_{add},  
\end{equation}
where $\bar{V}_{add}$ represents all additional terms that are bounded as $\epsilon\rightarrow0$. $\theta_1$, $\theta_2$, and $\theta_g$ correspond to the angles of the vortex sheets ($\gamma_{\gamma1}$, $\gamma_{\gamma2}$, and $\gamma_{T}$) in complex domain as they approaches the trailing edge. Now, we combine equation~(\ref{eq: VS_vel_TE3}) and $\text{Im}\{\bar{V}_{g}e^{i\theta_g}\} = 0$ (the complex form of equation~(\ref{eq: unsteady_Kutta})), and then divide both sides by the leading-order term, $\ln(\epsilon)$, to obtain $\gamma_g + \gamma_1 \cos(\theta_g - \theta_1) + \gamma_2 \cos(\theta_g - \theta_2) = 0$. We note that the term associated with $\bar{V}_{add}$ vanishes because $\ln(\epsilon)\rightarrow -\infty$ as $\epsilon\rightarrow0$. Furthermore, the no-slip boundary condition gives $\gamma_1 = u_{1+}$ and $\gamma_2 = -u_{2-}$. Together with the angle relations defined in figure~\ref{fig:TE-VS-schm}, the final equation takes the form
\begin{equation}
\label{eq: VS_gamma_TE}
 \gamma_g = - u_{g-} + u_{g+} = u_{1+} \cos{\Delta\theta_1} - u_{2-} \cos{\Delta\theta_2}.  
\end{equation}
For the case of a flat plate or a cusped trailing edge where both $\Delta \theta_1$ and $\Delta \theta_2$ are zero, equation~(\ref{eq: VS_gamma_TE}) is reduced to $\gamma_g = \gamma_1 + \gamma_2$, which is consistent with that given by \cite{JonesMA:03a}.

\subsection{Conservation of circulation}\label{sec:4-3}

Next, we proceed to analyze the change of circulation for the material volume $A_m$ in figure~\ref{fig:TE-schm}. The total change of the circulation within $A_m$ can be expressed as
\begin{equation}
\label{eq: tot_circulation}
 \begin{split}
 \frac{\mathrm{d}}{\mathrm{d}t} \int \omega \mathrm{d} A_m  = \frac{\mathrm{d}}{\mathrm{d}t} \oint_{\partial A_m} \boldsymbol{u} \cdot \hat{\boldsymbol{s}}_m \mathrm{d} s_m = \oint_{\partial A_m} \frac{\mathrm{d} \boldsymbol{u}}{\mathrm{d}t} \cdot \hat{\boldsymbol{s}}_m \mathrm{d} s_m + \oint_{\partial A_m} \frac{1}{2} \mathrm{d}(\boldsymbol{u} \cdot \boldsymbol{u}).
 \end{split}
\end{equation}
The left equation is based on the Green theorem. Since a vortex sheet corresponds to a velocity discontinuity, the Green theorem should be derived for a volume containing discontinuous surfaces as shown in appendix~\ref{sec:9}. We note that since the velocity derivative across a vortex sheet satisfies the Dirac delta function specified in equation~(\ref{eq: APX0_Int5}), the Green theorem should take its original form. In the right equation, since $\boldsymbol{u}$ is a Heaviside step function across any vortex sheet, it is single-valued throughout the entire domain. So the second term on the right hand side of equation~(\ref{eq: tot_circulation}) is equal to zero.

For an incompressible isotropic Newtonian fluid, the momentum equation on $\partial A_m$ can be expressed in the body-fixed reference frame as
\begin{equation}
\label{eq: navier-stokes}
 \frac{\mathrm{d} \boldsymbol{u}}{\mathrm{d} t} = \frac{1}{\rho} \left(-\nabla p + \nabla \cdot \bar{\bar{\tau}} \right) + \dot{\boldsymbol{u}}_{\Omega}, 
\end{equation}  
where $\bar{\bar{\tau}}$ is the shear stress tensor, and $\dot{\boldsymbol{u}}_{\Omega} = - 2 \boldsymbol{\Omega} \times \boldsymbol{u} - \boldsymbol{\Omega} \times (\boldsymbol{\Omega} \times \boldsymbol{r}) - \dot{\boldsymbol{U}}_b - \dot{\boldsymbol{\Omega}} \times \boldsymbol{r}$. Here, $\dot{\boldsymbol{U}}_b = - \dot{\boldsymbol{U}}$ represents the linear acceleration of the airfoil, where $\boldsymbol{U}$ is the translational background flow velocity at the infinity in the body-fixed reference frame. $\dot{\boldsymbol{\Omega}}$ is the angular acceleration of the airfoil and $\boldsymbol{r}$ is the position vector relative to the rotation center. Equation~(\ref{eq: navier-stokes}) can be plugged into equation~(\ref{eq: tot_circulation}) to give
\begin{equation}
\label{eq: tot_circulation1}
 \frac{\mathrm{d}}{\mathrm{d}t} \int \omega \mathrm{d} A_m = \oint_{\partial A_m} \left[\frac{1}{\rho} (-\nabla p + \nabla \cdot \bar{\bar{\tau}}) + \dot{\boldsymbol{u}}_{\Omega} \right] \cdot \hat{\boldsymbol{s}}_m \mathrm{d} s_m.
\end{equation}
This gives a general equation for the total change of circulation inside $A_m$. Now, we apply physical boundary conditions to simplify equation~(\ref{eq: tot_circulation1}). Since $S_{s1}$ and $S_{s2}$ correspond to fluid-solid interfaces that satisfy the no-slip boundary condition, we obtain $\mathrm{d} \boldsymbol{u}/\mathrm{d} t = 0$ in the body-fixed reference frame. Furthermore, the flow outside the vortex sheets are assumed to be inviscid, so we have $\nabla \cdot \bar{\bar{\tau}} = 0$ on the boundaries $\partial A_m - S_{s1} - S_{s2}$. Therefore, equation~(\ref{eq: tot_circulation1}) becomes
\begin{equation}
\label{eq: tot_circulation2}
\begin{split}
 \frac{\mathrm{d}}{\mathrm{d}t} \int \omega \mathrm{d} A_m  & = \int_{\partial A_m - S_{s1} - S_{s2}} \left(- \frac{\partial p}{\rho \partial s_m}  + \dot{\boldsymbol{u}}_{\Omega} \cdot \hat{\boldsymbol{s}}_m \right) \mathrm{d} s_m.
\end{split}
\end{equation}
Under condition (c) of Section~\ref{sec:4-1}, $\dot{\boldsymbol{u}}_{\Omega} \cdot \hat{\boldsymbol{s}}_m$ should be finite. Together with condition (a), the second integral of equation~(\ref{eq: tot_circulation2}) has the magnitude $\mathrm{O}(\epsilon_s)$. So equation~(\ref{eq: tot_circulation2}) has the simplified form
\begin{equation}
\label{eq: tot_circulation3}
\begin{split}
 \frac{\mathrm{d}}{\mathrm{d}t} \int \omega \mathrm{d} A_m  & = - \frac{1}{\rho} \int_{\partial A_m - S_{s1} - S_{s2}} \mathrm{d} p + \mathrm{O}(\epsilon_s).
\end{split}
\end{equation}
Let $p_1$ and $p_2$ be the pressure at the vertices of $A_m$ shown in figure~\ref{fig:TE-schm}, equation~(\ref{eq: tot_circulation3}) has the result $(p_2-p_1)/\rho + \mathrm{O}(\epsilon_s)$. Physically, pressure should be continuous at the trailing edge which means $p_1=p_2$ as $\epsilon_s \rightarrow 0$. In this case, equation~(\ref{eq: tot_circulation3}) becomes zero which returns the Kelvin's circulation theorem in the limit $\epsilon_s \rightarrow 0$. 

On the other hand, vorticity can be viewed as a material quantity moving with a fluid element. So the total change of circulation inside $A_m$ can be expressed using the Reynolds transport theorem as
\begin{equation}
\label{eq: vorticity_merge}
 \frac{\mathrm{d}}{\mathrm{d}t} \int \omega \mathrm{d} A_m = \int \frac{\partial \omega}{\partial t} \mathrm{d} A_m + \oint_{\partial A_m} \omega (\boldsymbol{u} \cdot \hat{\boldsymbol{n}}_m) \mathrm{d} s_m.
\end{equation}
The first term on the right hand side of equation~(\ref{eq: vorticity_merge}) can be written in the form 
\begin{equation}
\label{eq: vorticity_merge4}
 \int \frac{\partial \omega}{\partial t} \mathrm{d} A_m = \frac{\partial}{\partial t} \int \omega \mathrm{d} A_m = \frac{\partial \gamma_m}{\partial t} L_m,
\end{equation}
where $\gamma_m$ and $L_m$ represent the effective strength and length of the total vortex-sheet inside $A_m$, respectively. Again, $\partial \gamma_m/\partial t$ should be finite according to condition (c) of Section~\ref{sec:4-1}. With condition (a), $L_m \sim \mathrm{O}(\epsilon_s)$ so equation~(\ref{eq: vorticity_merge4}) also approaches zero in the limit $\epsilon_s \rightarrow 0$. Together with equation~(\ref{eq: tot_circulation2}) being zero as $\epsilon_s \rightarrow 0$, equation~(\ref{eq: vorticity_merge}) is reduced to
\begin{equation}
\label{eq: vorticity_merge5}
 \oint_{\partial A_m} \omega (\boldsymbol{u} \cdot \hat{\boldsymbol{n}}_m) \mathrm{d} s_m = 0.
\end{equation}     
Applying condition (b) of Section~\ref{sec:4-1} ($u_n = 0$ on $S_{f1}$ and $S_{f2}$) and $\omega = \partial u_n/\partial s_m$ on $S_{1}$, $S_{2}$, and $S_{g}$, equation~(\ref{eq: vorticity_merge5}) can be simplified as
\begin{equation}
\label{eq: vorticity_merge6}
 \int_{S_{1}+S_{g}+S_{2}}  u_n \mathrm{d} u_n = 0.
\end{equation} 
Plugging in the value of $u_n$ on $S_{1}$, $S_{2}$, and $S_{g}$, equation~(\ref{eq: vorticity_merge6}) becomes 
\begin{equation}
\label{eq: vorticity_merge7}
 \frac{1}{2}(u_{1+}^2 - u_{1-}^2) + \frac{1}{2}(u_{2+}^2 - u_{2-}^2) + \frac{1}{2}(u_{g+}^2 - u_{g-}^2) = 0.
\end{equation}
With the no-slip boundary condition, we obtain $u_{1-}=0$ and $u_{2+}=0$. For the free vortex sheet generated at the trailing edge, its strength and relative velocity satisfy $\gamma_{g} = -u_{g-}+u_{g+}$ and $u_{g} = (u_{g-}+u_{g+})/2$, respectively. Therefore, equation~(\ref{eq: vorticity_merge7}) can be combined with eq. 4.134 of \cite{WuJZ:06a} to give
\begin{equation}
\label{eq: vorticity_merge8}
 \frac{\partial \Gamma_{g}}{\partial t} = u_{g}\gamma_{g} = \frac{1}{2}(u_{2-}^2 - u_{1+}^2).
\end{equation} 
Note that similar formulations have also been obtained by \cite{SearsWR:56a,SearsWR:76a} and \cite{BasuBC:77a}, and it can be viewed as the differential form of the well-known Morino condition \citep{MorinoL:74a}. According to the above discussion, this condition determines the rate of circulation being shed from the trailing edge, and is valid for unsteady flows in the body-fixed reference frame. In addition, this condition is also consistent with Jones' condition for a flat plate (equation~(\ref{eq: FP_KC})).   


\section{Conservations of Mass and Momentum}\label{sec:5}   

In order to apply the conservation laws of mass and momentum to the merging process, we first provide further discussions on the insights of the bound vortex sheet introduced in Section~\ref{sec:3}. Generally, the non-penetration and no-slip boundary conditions are the physically correct conditions for fluid-solid interactions in most applications. While the Navier-Stokes equation allows for the matching of both normal and tangential velocity components between the fluid and the solid, the Euler equation allows only for the matching of the wall normal velocity component and it does not impose any constraints on the tangential velocity component. In order to remedy this for large Reynolds number flows, where the Euler equation is often accepted as a suitable model, we superimpose the Euler equations with a physical vortex sheet, $\gamma_{\gamma}$, as introduced in Section~\ref{sec:3} to satisfy the no-slip boundary condition. Therefore, $\gamma_{\gamma}$ actually represents the physical viscous shear layer at the fluid-solid interface, in the sense of preserving the tangential velocity jump or the circulation across the shear layer. As has been demonstrated in Section~\ref{sec:4-3}, the modeling of the physical vortex sheet allows one to perform calculations related to the formation of a free vortex sheet, especially the circulation relations. However, since the thickness and the velocity profile of a viscous shear layer are not resolved by a vortex sheet, the mass and momentum associated with the shear layer are not captured. Although this will not directly affect the solution of the original Euler equation, it would definitely cause unbalanced equations of mass and momentum within the vortex sheet, especially in the tangential direction, and thereby affecting the correct prediction of viscous shear force exerted on the shear layer in inviscid flows.
   
\subsection{A generalized sheet model for viscous shear layer}\label{sec:5-1}   

\begin{figure}
 \center{\includegraphics[width=0.7\textwidth]{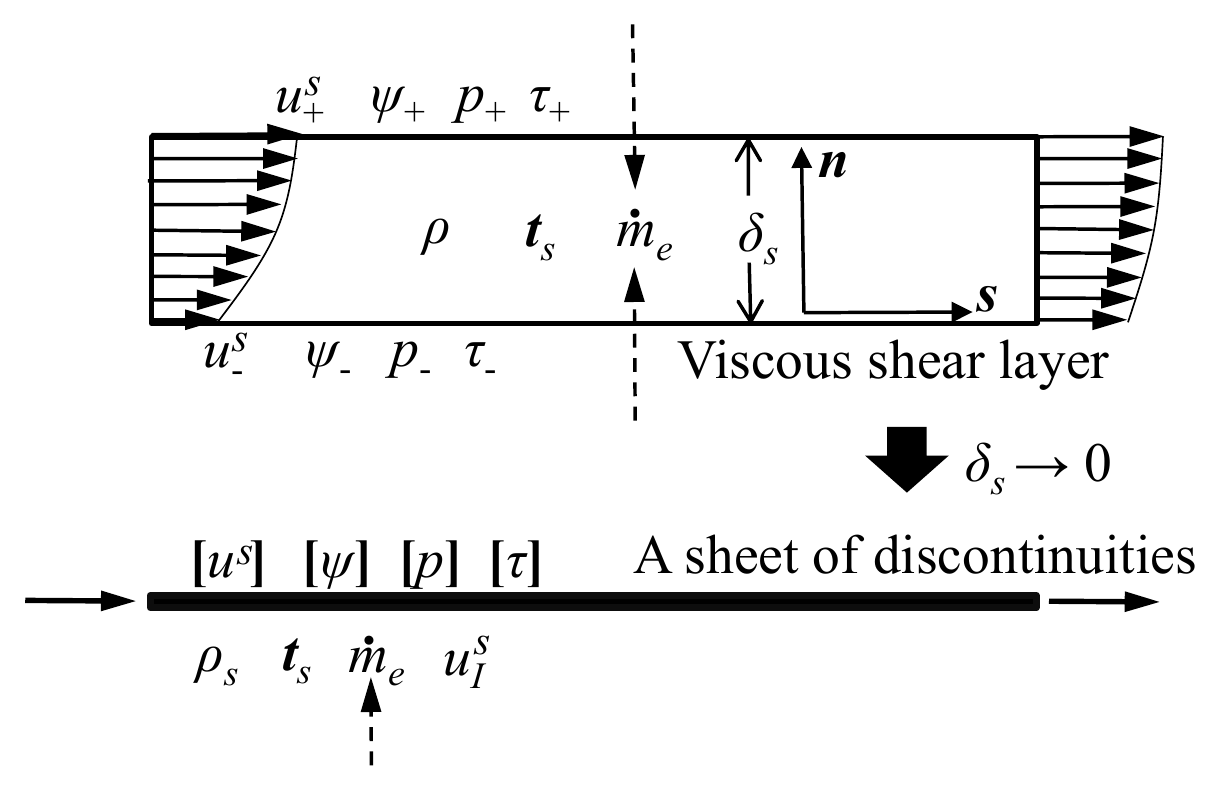}}
 \caption{\small A generalized sheet model to represent a viscous shear layer.}
 \label{fig:Gen-Sheet}
\end{figure}

In order to properly model the dynamics of a viscous shear layer, here we propose a generalized sheet model on top of the original vortex sheet where all relevant quantities or discontinuities associated with the viscous shear layer are superimposed. A schematic of this modeling approach is illustrated in figure~\ref{fig:Gen-Sheet}. As a first step, a sheet of discontinuity in the stream function $\psi$ is placed at the location of the original vortex sheet, so that $[\![\psi(s)]\!]$ is equal to the volumetric flow rate of the viscous shear layer in the form
\begin{equation}
\label{eq: IF_FR}
 [\![\psi]\!] = \int_0^{\delta_s} u^s \mathrm{d}n,
\end{equation} 
where $\delta_s$ is the thickness of the shear layer and $u^s$ is the tangential velocity component. Thus, the mass conservation for the new sheet can be written in the differential form
\begin{equation}
\label{eq: IF_mass}
 \frac{\mathrm{d}\rho_s}{\mathrm{d}t} = \rho \frac{\partial [\![\psi]\!]}{\partial s} - \dot{m}_e = 0,
\end{equation}
where $\dot{m}_e(s)$ is the per-unit-length mass entrainment associated with the shear layer and $\rho_s$ is the per-unit-length density defined as $\rho_s = \rho \delta_s$. 

To apply the momentum conservation law to a shear layer, we define a new discontinuity, $[\![\chi]\!]$, in analogy to $[\![\psi]\!]$ such that
\begin{equation}
\label{eq: IF_MomFlux}
 [\![\chi]\!] = \int_0^{\delta_s} (u^{s})^2 \mathrm{d}n.
\end{equation} 
Therefore, $[\![\chi]\!]$ represents the momentum flux associated with the generalized sheet. Furthermore, it is assumed that the new sheet has a characteristic velocity $\boldsymbol{u}_{I}(s) = u^s_{I}\hat{\boldsymbol{s}}$, satisfying $u^s_{I}=[\![\chi]\!]/[\![\psi]\!]$. In this way, the momentum flux of the shear layer is conserved. To further generalize the vortex sheet, we also superimpose a pressure jump, $[\![p(s)]\!]$, a shear stress jump, $[\![\tau(s)]\!]$, and a surface stress (tension) tensor, $\bar{\bar{T}}_s$, which is related to the surface stress $\boldsymbol{t}_s$ as $\boldsymbol{t}_s = \hat{\boldsymbol{s}} \cdot \bar{\bar{T}}_s$ in 2D. Now, applying the Reynolds transport theorem to a sheet element with a length of $\Delta s$, the momentum conservation can be expressed as
\begin{equation}
\label{eq: IF_momentum}
 \rho \frac{\mathrm{d} ([\![\psi]\!]\hat{\boldsymbol{s}})}{\mathrm{d}t} = \rho \frac{\partial ([\![\psi]\!]\hat{\boldsymbol{s}})}{\partial t} + \rho \frac{\partial ([\![\psi]\!]\boldsymbol{u}_{I})}{\partial s} - \dot{m}_e\boldsymbol{u}_e = -[\![p]\!]\hat{\boldsymbol{n}} + [\![\tau]\!]\hat{\boldsymbol{s}} + \nabla \cdot \bar{\bar{T}}_s,
\end{equation}                
where $\boldsymbol{u}_e$ is the velocity of the entrained fluid. We note that by assigning proper quantities and discontinuities this new sheet is capable of modeling the dynamics of a viscous shear layer at fluid-fluid or fluid-solid interface in single and multiple phase flows. 

Next, we investigate the application of equations~(\ref{eq: IF_mass}) and~(\ref{eq: IF_momentum}) for a special case, the physical vortex sheet $\gamma_{\gamma}$ at the surface of the airfoil. Firstly, equation~(\ref{eq: IF_mass}) can be integrated around the airfoil to give
\begin{equation}
\label{eq: IF_flux}
 \rho \sum [\![\psi_g]\!] - \int_0^{S_{B}}\dot{m}_e \mathrm{d}s = 0,
\end{equation} 
where $[\![\psi_g]\!]$ represents the stream function jump for each free vortex sheet being generated from $\gamma_{\gamma}$. Note that all quantities here are estimated in the body-fixed reference frame. Therefore, an additional term, $\rho \int_0^{\delta_s} \dot{\boldsymbol{u}}_{\Omega} \mathrm{d}n$, should be added to the right hand side of equation~(\ref{eq: IF_momentum}), with $\dot{\boldsymbol{u}}_{\Omega}$ being the same acceleration term as in equation~(\ref{eq: navier-stokes}). For simplicity, this term is ignored here by assuming $\delta_s \rightarrow 0$. We further assume the velocity of the entrained fluid to be identical to fluid side of the vortex sheet so $\boldsymbol{u}_e = \boldsymbol{u}_f - \boldsymbol{u}_b$, where $\boldsymbol{u}_{b} = u^s_{b}\hat{\boldsymbol{s}} + u^n_{b}\hat{\boldsymbol{n}}$. With the non-penetration boundary condition, we have $\boldsymbol{u}_{f} = u^s_{f}\hat{\boldsymbol{s}} + u^n_{b}\hat{\boldsymbol{n}}$ and $\gamma_{\gamma}=u^s_{f}-u^s_{b}$. Neglecting surface tension and plugging in equation~(\ref{eq: IF_mass}), equation~(\ref{eq: IF_momentum}) can be expressed in the $\hat{\boldsymbol{s}}$ and $\hat{\boldsymbol{n}}$ directions as
\begin{align}
\label{eq: IF_momentum_s}
 \rho \left(\frac{\partial [\![\psi]\!]}{\mathrm{d}t} + [\![\psi]\!]\frac{\partial u^s_{I}}{\partial s} +  \frac{\partial [\![\psi]\!]}{\partial s}(u^s_{I}-\gamma_{\gamma}) \right) \hat{\boldsymbol{s}} &= [\![\tau]\!]\hat{\boldsymbol{s}},\\
\label{eq: IF_momentum_n}
 \boldsymbol{0} &= [\![p]\!] \hat{\boldsymbol{n}}.
\end{align}   
Equation~(\ref{eq: IF_momentum_n}) is still consistent with previous studies \citep{Saffman:92a,WuJZ:06a} that pressure is continuous across a vortex sheet. This means that the generalized sheet model does not affect the force balance in the normal direction of the sheet. In this sense, equation~(\ref{eq:AF_force}) still captures the total force contributed from the pressure term. Now, we further integrate equation~(\ref{eq: IF_momentum_s}) around the airfoil to obtain
\begin{equation}
\label{eq: IF_shear}
 [\![\boldsymbol{f}_{\tau}]\!] = \rho \int_0^{S_{B}} \left(\frac{\partial [\![\psi]\!]}{\mathrm{d}t} + [\![\psi]\!]\frac{\partial u^s_{I}}{\partial s} + \frac{\partial [\![\psi]\!]}{\partial s}(u^s_{I}-\gamma_{\gamma}) \right) \hat{\boldsymbol{s}} \mathrm{d}s + \rho\sum [\![\psi_g]\!]\boldsymbol{u}^*_{g},
\end{equation}   
where $[\![\boldsymbol{f}_{\tau}]\!]$ is the jump of the total shear force between the fluid and solid sides of the vortex sheet around the airfoil. $\boldsymbol{u}^*_{g}$ represents the momentum-based characteristic velocity associated with each free vortex sheet coming off from the airfoil; a special case of trailing-edge vortex-sheet formation is provided in Section~\ref{sec:5-2}. Since the fluid side of $\gamma_{\gamma}$ is a free shear surface with zero shear stress, $[\![\boldsymbol{f}_{\tau}]\!]$ is actually the unsteady viscous drag exerted by the solid body, which is not captured by equation~(\ref{eq:AF_force}). Similar to that reported by \cite{WuJZ:15a}, the term $[\![\psi]\!]$ in this study is also the core parameter in drag generation, while here the calculation is performed for the unsteady case. Last, we emphasize that this generalized sheet model enables the estimation of viscous force based on inviscid flow, although several relevant \emph{global} quantities and discontinuities around the airfoil have to be modeled or known a priori.

Before applying this modeling approach to the merging process at the trailing edge, here we briefly summarize the generalized sheet model. In seeking reduced flow models and force calculations instead of solving the Navier-Stokes equation, inviscid flow and vortex models have been adopted to account for the effect of viscous regions without resolving the actual distribution of vorticity. A main idea was to model shear layers and wake vortices by superimposing bound and free vortex sheets at the equation level. In this study, we propose to further superimpose necessary \emph{global} quantities and discontinuities at the location of the original vortex sheet at the solution level for improved force calculation and accurate prediction of vortex-sheet formation, as summarized in Table~\ref{tab:Mdl_sum}.
\begin{table}
\begin{center}
\begin{tabular}{ccLL}
\vspace{3mm}
 Name & Symbol & Feature & Captured global quantity\\ \vspace{2mm} 
Free vortex sheet & $\gamma$ & Free shear surfaces at both sides & Circulation per unit-length of wake shear layer \\ \vspace{2mm}
Bound vortex sheet & $\gamma_{\gamma}$ & Free shear surface at one side; no-slip at the other side & Circulation per unit-length of body shear layer \\ \vspace{2mm}
Mass-flux sheet & $[\![\psi]\!]$ & Captures the entrainment & Mass flux\\ \vspace{2mm}
Momentum-flux sheet & $[\![\chi]\!]$ & Enables the analysis of momentum transportation & Momentum flux\\ \vspace{2mm}
Energy-flux sheet & $[\![\lambda]\!]$ & Enables the analysis of energy dissipation & Flux of kinetic energy\\ \vspace{2mm}
Stress sheet & $[\![\sigma]\!]$ & Enables the force analysis, especially the viscous force & Jumps of pressure, shear stress, or surface stress\\ \vspace{2mm}

\end{tabular}
\caption{\small{A summary of the sheet models for a viscous shear layer. $[\![\lambda]\!]$ is defined as $[\![\lambda]\!] = \int_0^{\delta_s} (u^{s})^3 \mathrm{d}n$}}.
\label{tab:Mdl_sum}
\end{center}
\end{table}

\subsection{A special case: the finite-angle trailing edge}\label{sec:5-2} 

With the proposed generalized sheet model, we now proceed to apply mass and momentum conservations to the control volume $A_m$ of figure~\ref{fig:TE-schm} to obtain further information about the forming vortex sheet at the trailing edge. Here, it is important to note that the mass and momentum equations should be performed for the entire $A_m$ rather than just the triple-joint sheet structure, because the inviscid flow outside the sheet system also plays an essential part in dictating the flow regime near the trailing edge. In fact, for the high-Reynolds number case where the mass and momentum contributions from the viscous shear layer become negligible, the direction of the trailing-edge streamline should be solely governed by the inviscid flow. For the mass conservation, since there is no mass flux across the trailing-edge streamline, the mass conversation for $A_m$ can be written separately for $A_{m1}$ and $A_{m2}$ in the form
\begin{align}
\label{eq: VS_masscon1}
 \frac{\mathrm{d}}{\mathrm{d}t} \int_{A_{m1}} \rho \mathrm{d} A = \int_{A_{m1}} \frac{\partial \rho}{\partial t} \mathrm{d} A + \oint_{\partial A_{m1}} \rho (\boldsymbol{u} \cdot \hat{\boldsymbol{n}}_m) \mathrm{d} s_m = 0,\\
\label{eq: VS_masscon2}
 \frac{\mathrm{d}}{\mathrm{d}t} \int_{A_{m2}} \rho \mathrm{d} A = \int_{A_{m2}} \frac{\partial \rho}{\partial t} \mathrm{d} A + \oint_{\partial A_{m2}} \rho (\boldsymbol{u} \cdot \hat{\boldsymbol{n}}_m) \mathrm{d} s_m = 0.
\end{align}
With the generalized sheet model proposed in Section~\ref{sec:5-1}, the momentum conservation for $A_{m}$ can be expressed in the form
\begin{equation}
\label{eq: VS_momentumcon1}
\begin{split}
 \frac{\mathrm{d}}{\mathrm{d}t} \int \rho \boldsymbol{u} \mathrm{d} A_m  &= \int \frac{\partial (\rho \boldsymbol{u})}{\partial t} \mathrm{d} A_{m} + \oint_{\partial A_{m}} \rho \boldsymbol{u} (\boldsymbol{u} \cdot \hat{\boldsymbol{n}}_m) \mathrm{d} s_m\\  
 &= \int \left(-\nabla p + \nabla \cdot \bar{\bar{\tau}} + \rho \dot{\boldsymbol{u}}_{\Omega}  \right) \mathrm{d} A_{m-} + \int_{S_{\gamma 1}+S_{\gamma 2}+S_{\gamma g}} ([\![ \boldsymbol{\tau} ]\!] + \nabla \cdot \bar{\bar{T}}_s) \mathrm{d} s,
\end{split}
\end{equation}
where $S_{\gamma 1}$, $S_{\gamma 2}$, and $S_{\gamma 3}$ correspond to the vortex sheets in $A_{m}$; $A_{m-}$ denotes the volume of $A_{m}$ excluding $S_{\gamma 1}$, $S_{\gamma 2}$, and $S_{\gamma 3}$. Again, $\boldsymbol{\tau} = \hat{\boldsymbol{n}} \cdot \bar{\bar{\tau}}$ is the shear stress. $\bar{\bar{T}}_s$ satisfies $\boldsymbol{t}_s = \hat{\boldsymbol{s}} \cdot \bar{\bar{T}}_s$ where $\boldsymbol{t}_s$ is the surface stress. The first term in the second equation is from equation~(\ref{eq: navier-stokes}), whereas the second term is from equation~(\ref{eq: IF_momentum}) with pressure jump across the sheet being zero.    

Next, we simplify equations~(\ref{eq: VS_masscon1}), ~(\ref{eq: VS_masscon2}), and~(\ref{eq: VS_momentumcon1}) for the specific problem of vortex-sheet formation at a sharp edge. Considering the infinitesimal size of the control volume normal to the vortex sheets, any variation of the velocity over $S_{1}$, $S_{2}$, $S_{g-}$, and $S_{g+}$ is neglected, and equations~(\ref{eq: VS_masscon1}) and~(\ref{eq: VS_masscon2}) can be transformed into the form  
\begin{align}
\label{eq: VS_mass1}
u_{1+}S_{1} + [\![ \psi_1 ]\!] + u_{g-}S_{g-} + [\![ \psi_{g-} ]\!] = 0,\\
\label{eq: VS_mass2}
u_{2-}S_{2} + [\![ \psi_2 ]\!] + u_{g+}S_{g+} + [\![ \psi_{g+} ]\!] = 0.
\end{align}
The jump of the stream function is applied here to account for the mass flux associated with a vortex sheet, as defined in Section~\ref{sec:4}. We further note that the mass flux associated with the forming vortex sheet is divided into $[\![ \psi_{g-} ]\!]$ and $[\![ \psi_{g-} ]\!]$ by the trailing-edge streamline. For equation~(\ref{eq: VS_momentumcon1}), we apply conditions (a) and (c) of Section~\ref{sec:4-1} to argue that the integral associated with $\partial(\rho \boldsymbol{u})/\partial t$ has the magnitude $\mathrm{O}(\epsilon_s^2)$. With $\nabla \cdot \bar{\bar{\tau}}=0$ for region outside the vortex sheets, equation~(\ref{eq: VS_momentumcon1}) has the form
\begin{equation}
\label{eq: VS_momentum0}
 \oint_{\partial A_{m}} \boldsymbol{u} (\boldsymbol{u} \cdot \hat{\boldsymbol{n}}_m) \mathrm{d} s_m + \mathrm{O}(\epsilon_s^2) = \int \left(-\frac{\nabla p}{\rho} + \dot{\boldsymbol{u}}_{\Omega}  \right) \mathrm{d} A_{m-} + \frac{1}{\rho}\int_{S_{\gamma 1}+S_{\gamma 2}+S_{\gamma g}} ([\![ \boldsymbol{\tau} ]\!] + \nabla \cdot \bar{\bar{T}}_s) \mathrm{d} s.
\end{equation}
Physically, $p$ is continuous so $\nabla p$ should be finite. Together with the boundedness of $\dot{\boldsymbol{u}}_{\Omega}$, the first integral on the right hand side of equation~(\ref{eq: VS_momentum0}) has the magnitude $\mathrm{O}(\epsilon_s^2)$. Therefore, equation~(\ref{eq: VS_momentum0}) is further reduced to
\begin{equation}
\label{eq: VS_momentum1}
 \oint_{\partial A_{m}} \boldsymbol{u} (\boldsymbol{u} \cdot \hat{\boldsymbol{n}}_m) \mathrm{d} s_m + \mathrm{O}(\epsilon_s^2) =  \frac{1}{\rho}\int_{S_{\gamma 1}+S_{\gamma 2}+S_{\gamma g}} ([\![ \boldsymbol{\tau} ]\!] + \nabla \cdot \bar{\bar{T}}_s) \mathrm{d} s.
\end{equation} 
In this study, there is no surface tension so $\nabla \cdot \bar{\bar{T}}_s = 0$. $S_{\gamma g}$ corresponds to the free vortex sheet which physically means $[\![ \boldsymbol{\tau} ]\!] = 0$. Furthermore, $[\![ \boldsymbol{\tau} ]\!]$ is finite on the bound vortex sheets $S_{\gamma 1}$ and $S_{\gamma 2}$ according to equation~(\ref{eq: IF_momentum_s}). Together with condition (a) of Section~\ref{sec:4-1}, the right hand side of equation~(\ref{eq: VS_momentum1}) has the magnitude $\mathrm{o}(\epsilon_s)$. Now, applying the velocity boundary conditions of $\partial A_{m}$, the momentum balance can be written in the $\hat{\boldsymbol{s}}_{g}$ and $\hat{\boldsymbol{n}}_{g}$ directions, respectively, in the form
\begin{equation}
\begin{split}
\label{eq: VS_momentum_s}
 & (u_{1+}^2 S_{1} + [\![ \psi_1 ]\!]u^*_{1+}) \cos{\Delta \theta_1} + (u_{2-}^2 S_{2} + [\![ \psi_2 ]\!]u^*_{2+}) \cos{\Delta \theta_2}\\
 &= u_{g-}^2S_{g-} + [\![ \psi_{g-} ]\!]u^*_{g-} + u_{g+}^2S_{g+} + [\![ \psi_{g+} ]\!]u^*_{g+} + O(\epsilon_s^2) + o(\epsilon_s),\\
\end{split}
\end{equation}
\begin{equation}
\begin{split}
\label{eq: VS_momentum_n}
 & (u_{1+}^2 S_{1} + [\![ \psi_1 ]\!]u^*_{1+}) \sin{\Delta \theta_1}\\
 &= (u_{2-}^2 S_{2} + [\![ \psi_2 ]\!]u^*_{2+}) \sin{\Delta \theta_2} + O(\epsilon_s^2) + o(\epsilon_s),
\end{split}
\end{equation}
where the superscript $^*$ denotes the characteristic velocity scale based on the momentum flux of a vortex sheet, similar to $u^s_{I}$ of Section~\ref{sec:5-1}. 

In general, the $[\![ \psi ]\!]$ and $u^*$ terms need to be given or solved as discussed in Section~\ref{sec:5-1}. In the current study, for the high-Reynolds number case, the mass and momentum associated with the vortex sheet are neglected as a first approximation. So the $[\![ \psi ]\!]$ and $u^*$ terms are assumed to be zero to solve equations~(\ref{eq: VS_mass1}), (\ref{eq: VS_mass2}), (\ref{eq: VS_momentum_s}), and (\ref{eq: VS_momentum_n}). The terms $\mathrm{O}(\epsilon_s^2)$ and $\mathrm{o}(\epsilon_s)$ can be neglected in the limit $\epsilon_s \rightarrow 0$. It is followed from equation~(\ref{eq: VS_momentum_n}) that $\Delta \theta_1 \cdot \Delta \theta_2 \geq 0$. Given
\begin{equation}
\label{eq: VS_angle0}
\Delta \theta_1+\Delta \theta_2 = \Delta \theta_0,
\end{equation} 
it yields $\Delta \theta_1, \Delta \theta_2 \geq 0$ which means that the direction of the forming vortex sheet should vary between the two tangents of the trailing edge surfaces. Now, we combine equations~(\ref{eq: VS_mass1}),~(\ref{eq: VS_mass2}),~(\ref{eq: VS_momentum_s}), and ~(\ref{eq: VS_momentum_n}), and cancel $S_{1}$, $S_{2}$, $S_{g-}$, and $S_{g+}$ to obtain
\begin{equation}
\label{eq: VS_momentum}
u_{1+}u_{2-} \sin{\Delta\theta_0} - u_{1+}u_{g+}\sin{\Delta\theta_1} - u_{2-}u_{g-}\sin{\Delta\theta_2} = 0.
\end{equation}
This gives the final equation of the mass and momentum balance at the trailing edge for high-Reynolds number cases. We note from this derivation that the surfaces of $A_m$ ($S_{1}$, $S_{2}$, $S_{g-}$, and $S_{g+}$) are not arbitrary and their relative dimensions are dictated by the mass and momentum conservation laws.
  
\section{Final Result and Discussions}\label{sec:6}

In this section, we first apply the conditions obtained in sections~\ref{sec:4} and \ref{sec:5} to determine the formation of the trailing-edge vortex sheet. Equations~(\ref{eq: vorticity_merge7}) and~(\ref{eq: VS_gamma_TE}) can now be combined to derive $u_{g-}$ and $u_{g+}$ in terms of $u_{1+}$, $u_{2-}$, $\Delta\theta_1$, and $\Delta\theta_2$, and the results can be plugged into equation~(\ref{eq: VS_momentum}) to obtain
\begin{equation}
\begin{dcases}
\label{eq: VS_combine1}
 (u_{1+} \sin{\Delta\theta_1} + u_{2-} \sin{\Delta\theta_2})^2 (u_{1+} \sin{\Delta\theta_1} - u_{2-} \sin{\Delta\theta_2}) & = 0 \hspace{3mm} \text{for} \hspace{3mm} u_{1+} \neq u_{2-},\\  
 -\cos{\Delta\theta_1} + \cos{\Delta\theta_2} & = 0 \hspace{3mm} \text{for} \hspace{3mm} u_{1+} = u_{2-}. 
\end{dcases}
\end{equation}
Since we have concluded from the momentum conservation in Section~\ref{sec:5-2} that $\Delta \theta_1, \Delta \theta_2 \geq 0$, equation~(\ref{eq: VS_combine1}) can be simplified as
\begin{equation}
\label{eq: VS_combine3}
 u_{1+} \sin{\Delta\theta_1} - u_{2-} \sin{\Delta\theta_2} = 0.  
\end{equation}
Again, due to $\Delta \theta_1, \Delta \theta_2 \geq 0$ and $0 \leq \Delta \theta_0 < \pi$, this equation indicates that $u_{1+}$ and $u_{2-}$ cannot take different signs. In the current study of vortex shedding ($u_{g-}, u_{g+} \geq 0$), this further indicates $u_{1+}, u_{2-} \leq 0$ which means no backward flow. Finally, combining equations~(\ref{eq: VS_angle0}) and~(\ref{eq: VS_combine3}) yields        
\begin{equation}
\label{eq: VS_combine4}
\begin{dcases}
 \Delta\theta_1 = \cos^{-1}\left(\frac{u_{1+}^2+u_3^2-u_{2-}^2}{2u_{1+}u_3}\right), \hspace{1mm} \Delta\theta_2 = \cos^{-1}\left(\frac{u_{2-}^2+u_3^2-u_{1+}^2}{2u_{2-}u_3}\right) \hspace{3mm} & \text{for} \hspace{3mm} u_{1+},u_{2-} < 0,\\
 \Delta\theta_1 = 0, \hspace{1mm} \Delta\theta_2 = \Delta\theta_0 & \text{for} \hspace{3mm} u_{2-} = 0,\\ 
 \Delta\theta_1 = \Delta\theta_0, \hspace{1mm} \Delta\theta_1 = 0 & \text{for} \hspace{3mm} u_{1+} = 0, 
\end{dcases}
\end{equation}
where $u_3 = -\sqrt{u_{1+}^2+u_{2-}^2+2u_{1+}u_{2-}\cos{\Delta\theta_0}}$. Therefore, equation~(\ref{eq: VS_combine4}) determines the direction of the forming vortex sheet at the trailing edge, and the result can be further combined with equations~(\ref{eq: vorticity_merge8}) and~(\ref{eq: VS_gamma_TE}) to obtain the analytical vortex-sheet strength ($\gamma_g$) and relative velocity ($u_g$).

The classical Kutta condition requires the rear stagnation streamline of an airfoil to be attached to the sharp trailing edge. Physically, this means that flow cannot turn around the sharp edge. For steady flow at the trailing edge, \cite{PolingDR:86a} has summerized a number of conditions that are equivalent to this condition:\\
1) Continuous pressure at the trailing edge.\\
2) The velocity at the trailing edge is finite or zero.\\
3) The shedding of vorticity vanishes ($\dot{\Gamma}_g = 0$).\\
4) The stagnation streamline bisects the wedge angle of the trailing edge ($\Delta\theta_1=\Delta\theta_2$).

For unsteady flow at the trailing edge, only condition 1 is valid according to \cite{BasuBC:77a} and \cite{PolingDR:86a}. The difference for an unsteady trailing-edge flow lies in the ambiguity of the direction of the stagnation streamline line. \cite{GiesingJP:69a} and \cite{MaskellEC:71a} have proposed that either $\Delta\theta_1=0$ or $\Delta\theta_2=0$ should be satisfied at the trailing edge. Although \cite{PolingDR:86a} have provided experimental support for this model when $\dot{\Gamma}_g$ is large, they also pointed out a serious flaw that the Giesing-Maskell model does not approach the steady solution (condition 4) as $\dot{\Gamma}_g \rightarrow 0$. \cite{PolingDR:86a} further confirmed this flaw as they observed a smooth change between the senarios of $\Delta\theta_1=0$ and $\Delta\theta_2=0$ when $\dot{\Gamma}_g$ approaches zero. 
\begin{figure}
 \center{\includegraphics[width=0.6\textwidth]{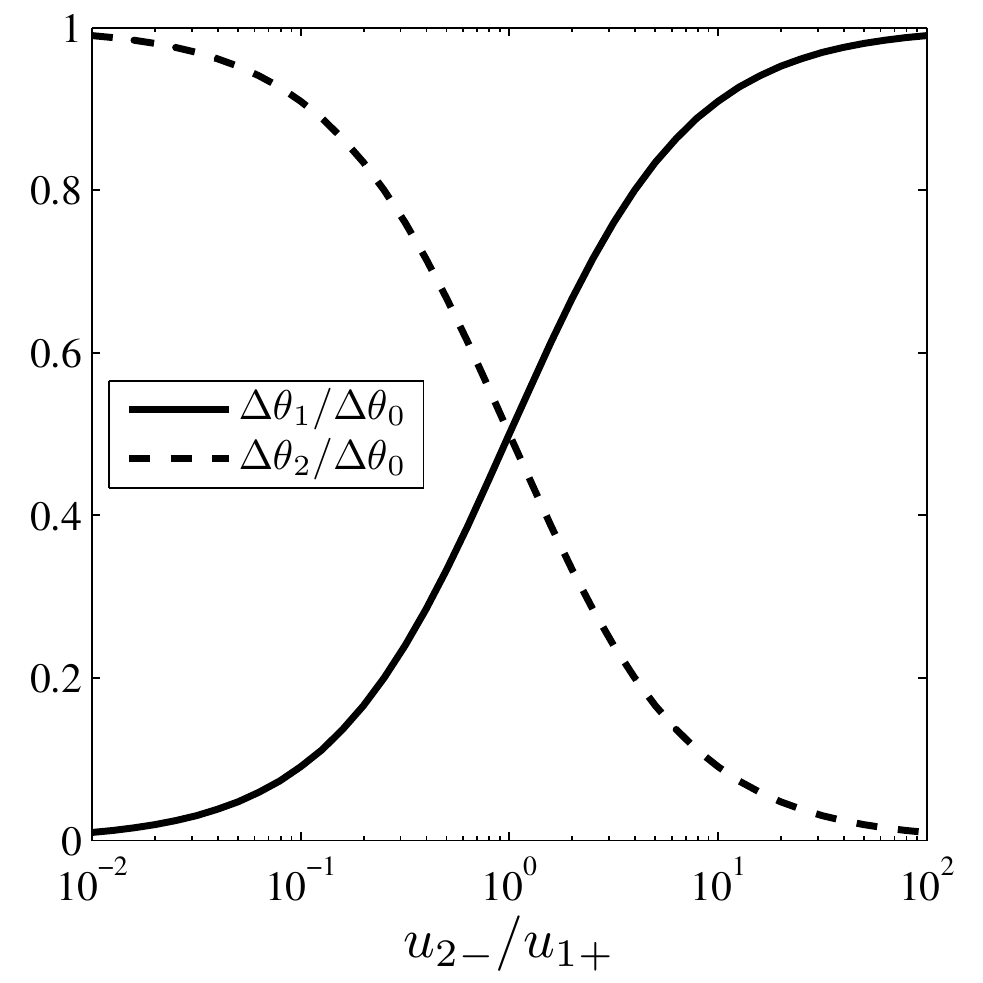}}
 \caption{\small The angle of the stagnation streamline (or the forming vortex sheet) vs. the ratio between $u_{2-}$ and $u_{1+}$.}
 \label{fig:angle-u}
\end{figure}

The current model provides a compelling explanation for the flaw of the Giesing-Maskell model, as we have analytically derived in equation~(\ref{eq: VS_combine4}) the relationship between the angle of the stagnation streamline (or forming vortex sheet) and the flow velocities at both sides of the trailing edge. The result of equation~(\ref{eq: VS_combine4}) can be interpreted by figure~\ref{fig:angle-u}, where $\Delta \theta_1$ and $\Delta \theta_2$ vary between 0 and $\Delta \theta_0$, and are solely determined by $u_{2-}/u_{1+}$. We note that $\Delta\theta_1=0$ or $\Delta\theta_2=0$ can be obtained as $u_{2-}=0$ or $u_{1+}=0$, respectively. This indicates that the Giesing-Maskell model actually corresponds to the two limiting cases of the current model. We also note that condition 4 of the steady solution can be recovered at $u_{2-}/u_{1+}=1$. Most importantly, the continuous transition between the Giesing-Maskell model and the steady solution is fully captured as $u_{2-}/u_{1+}$ varies between 0 and $\infty$. Therefore, we believe that the flaw of the Giesing-Maskell model is resulted from the non-physical assumption that the potential flow on either side of the trailing edge has to be stagnant on all occasions. In fact, this stagnation assumption could be true on the suction side of the trailing edge if the preceding flow has already separated. However, if the flow remains attached on both side of the trailing edge, the flow being stagnant on either side of the trailing edge is not justified.   
\begin{figure}
 \center{\includegraphics[width=1.0\textwidth]{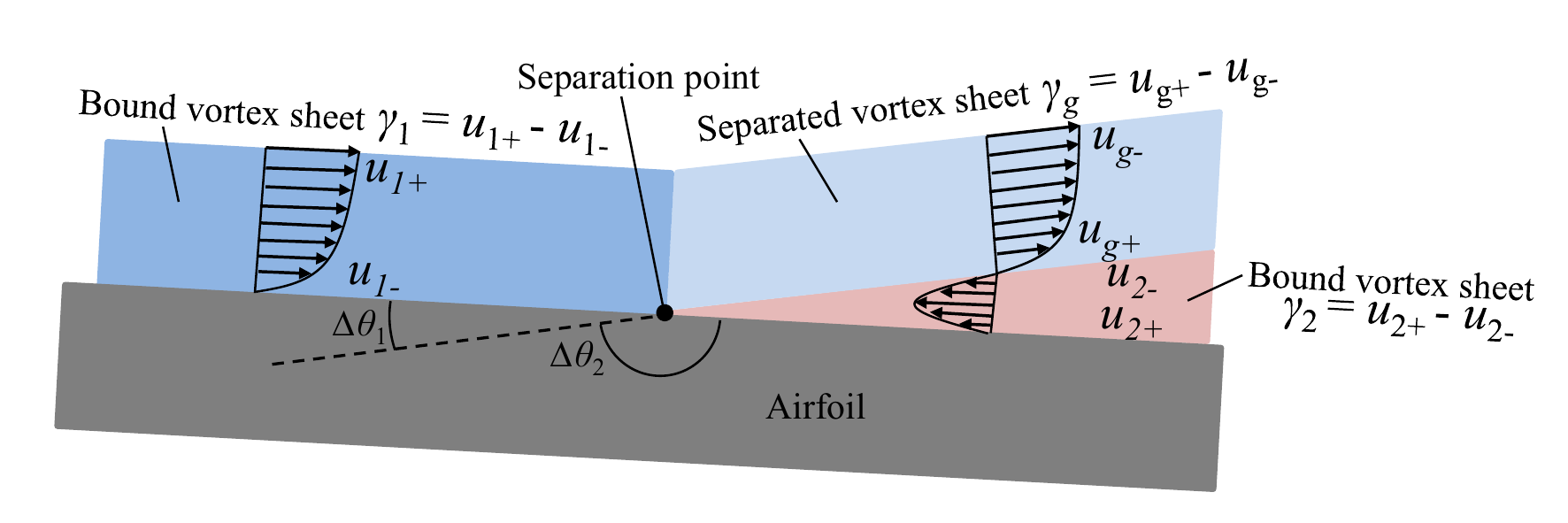}}
 \caption{\small The structure of viscous sheer layers and the corresponding vortex sheets near a flow separation point on a smooth surface. It is important to note that $u_{2+}=u_{2-}=0$ and $\gamma_2=0$.}
 \label{fig:LE-schm}
\end{figure}

The current model for the trailing-edge vortex sheet is based on conservation laws and the unsteady Kutta condition which only requires a continuous pressure distribution. In this sense, there should not be any fundamental difference for the formation of a vortex sheet due to flow separation on a smooth surface. Thus, we further propose to extend this model to deciding the formation of a leading-edge vortex sheet. For this purpose, the separated vortex sheet can be viewed as being generated due to the merging of the two bound vortex sheets at both sides of the separation point. Considering the actual viscous shear layers near a separation point as shown in figure~\ref{fig:LE-schm}, the downstream-side shear layer consists of a reverse-flow layer and a separated-flow layer. In the vortex-sheet limit, the reverse-flow layer becomes the bound vortex sheet while the separated-flow layer becomes the separated vortex sheet. Apparently, the velocities at both sides of the reverse-flow layer are zero ($u_{2+}=u_{2-}=0$), meaning the corresponding bound vortex-sheet strength is zero ($\gamma_2=0$) near the separation point. Based on the above discussions, we can attribute the formation of the separated vortex sheet to the scenario of the Giesing-Maskell model. Because $\Delta\theta_0=\pi$ for a smooth surface, we immediately obtain $\Delta\theta_1=0$ and $\Delta\theta_2=\pi$, which means that the forming vortex sheet from a separation point of a smooth surface should be tangential to the surface. Finally, applying equations~(\ref{eq: vorticity_merge8}) and~(\ref{eq: VS_gamma_TE}) gives the strength and velocity of the forming vortex sheet, which are actually equal to the values of its upstream bound vortex sheet ($\gamma_g=\gamma_1$ and $u_g=u_{1+}/2$). 

At this point, we have accomplished the task of analytically deciding the direction ($\Delta\theta_1$ and $\Delta\theta_2$), strength ($\gamma_{g}$), and relative velocity ($u_{g}$) of free vortex sheets formed at both the sharp trailing edge and the leading-edge separation point. 


\section{Simulations and validations}\label{sec:7}

To verify the unsteady flow model together with the vortex-sheet formation conditions for a 2D airfoil, this section will simulate different airfoils in steady and unsteady background flows, and then compare the results with experimental data or empirical models. Here, we note that the formation of the leading-edge vortex sheet at large angle-of-attack (AoA) requires predicting the leading-edge separation point, which could be a topic of a future investigation. To this point, the following assumes no flow separation near the leading edge and only deals with vortex shedding at the trailing edge. For this reason, the simulations of this study are limited to small-to-medium AoA regimes, where the flow might be considered to remain attached without loosing much accuracy.
\begin{figure}
\begin{center}

\begin{minipage}{0.6\linewidth} \begin{center}  
\includegraphics[width=.99\linewidth]{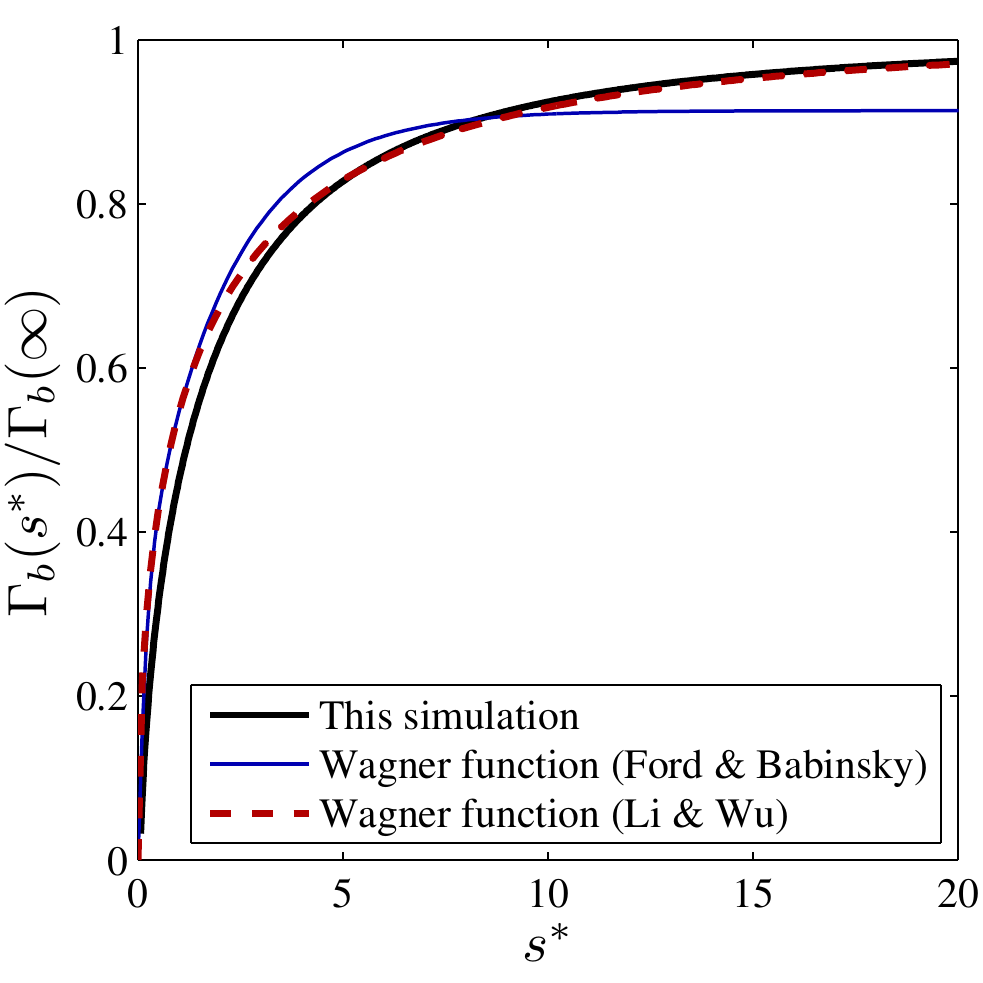}
\end{center} \end{minipage}\\

\caption{\small Non-dimensional bound circulation vs. non-dimensional distance traveled. The Wagner functions for circulation are provided by \cite{BabinskyH:13a} and \cite{WuZ:15a}, respectively.} 
\label{fig:wagner_comp}

\end{center}
\end{figure}
\begin{figure}
\begin{center}

\begin{minipage}{0.49\linewidth} \begin{center}  
\includegraphics[width=0.98\linewidth]{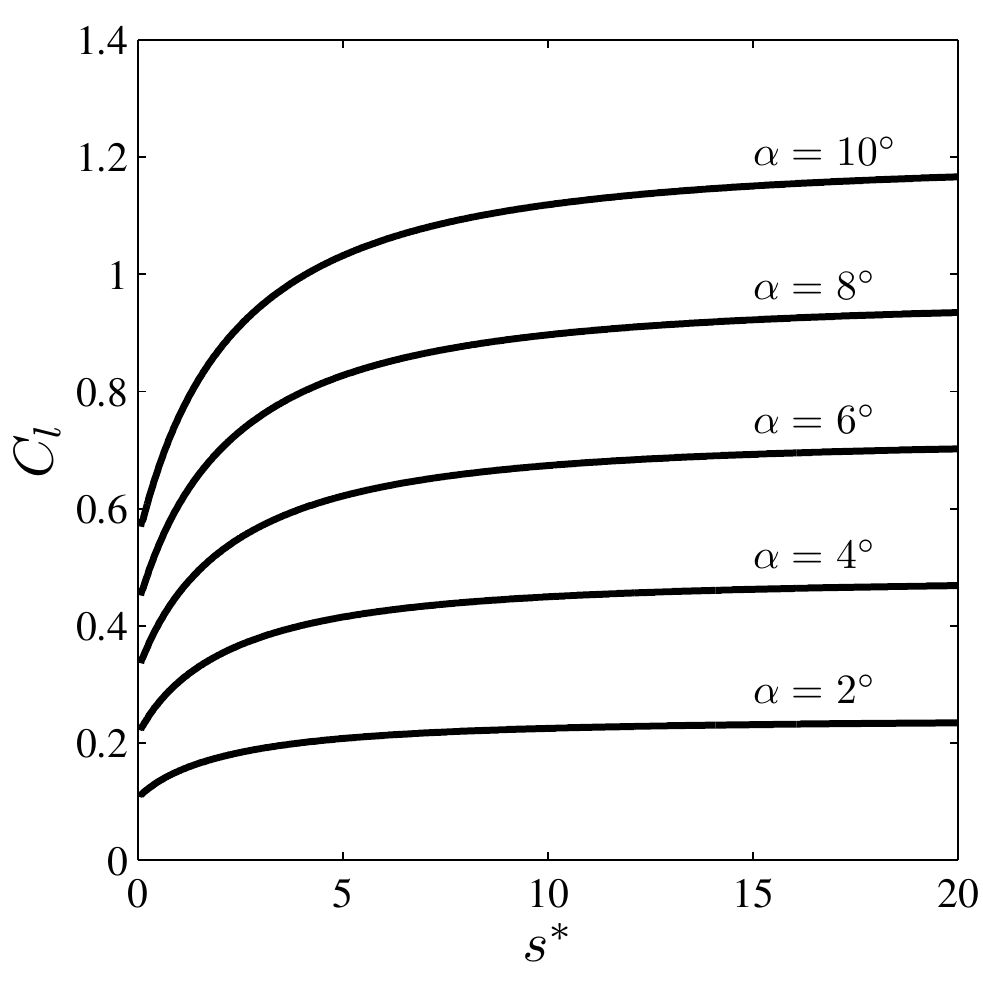}
\end{center} \end{minipage}
\begin{minipage}{0.49\linewidth} \begin{center}  
\includegraphics[width=1\linewidth]{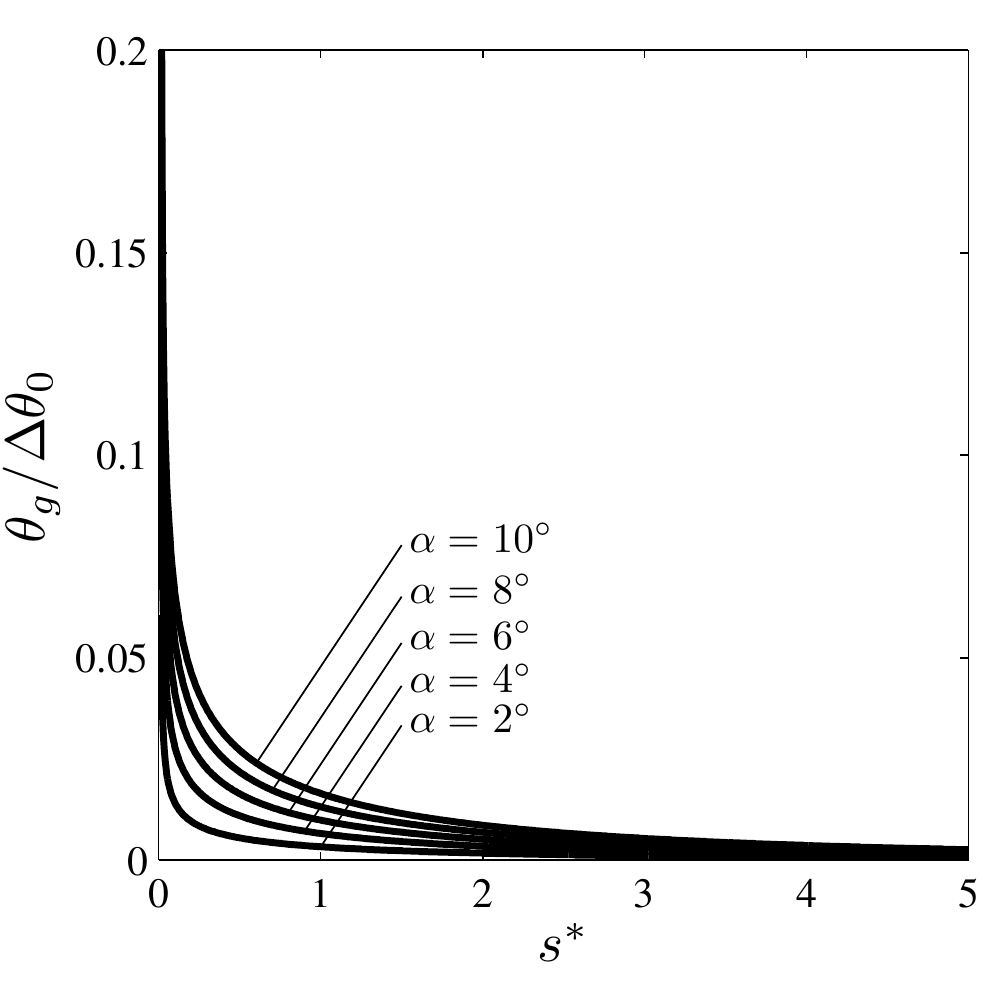}
\end{center} \end{minipage}\\

\begin{minipage}{0.49\linewidth} \begin{center} (a) \end{center} \end{minipage}
\begin{minipage}{0.49\linewidth} \begin{center} (b) \end{center} \end{minipage}

\caption{\small (a) Lift coefficient ($C_l$) vs. non-dimensional distance traveled ($s^*$) for a NACA 0012 airfoil at various angles of attack. (b) The variation of the angle of the trailing-edge vortex sheet. $\theta_g=0$ corresponds to the bisector of the finite-angle trailing edge. $\theta_g$ varies between $-\Delta \theta_0/2$ and $\Delta \theta_0/2$ that correspond to the two tangents of the trailing edge, respectively.} 
\label{fig:lift_steady1}

\end{center}
\end{figure}
\begin{figure}
\begin{center}

\begin{minipage}{0.49\linewidth} \begin{center}  
\includegraphics[width=0.99\linewidth]{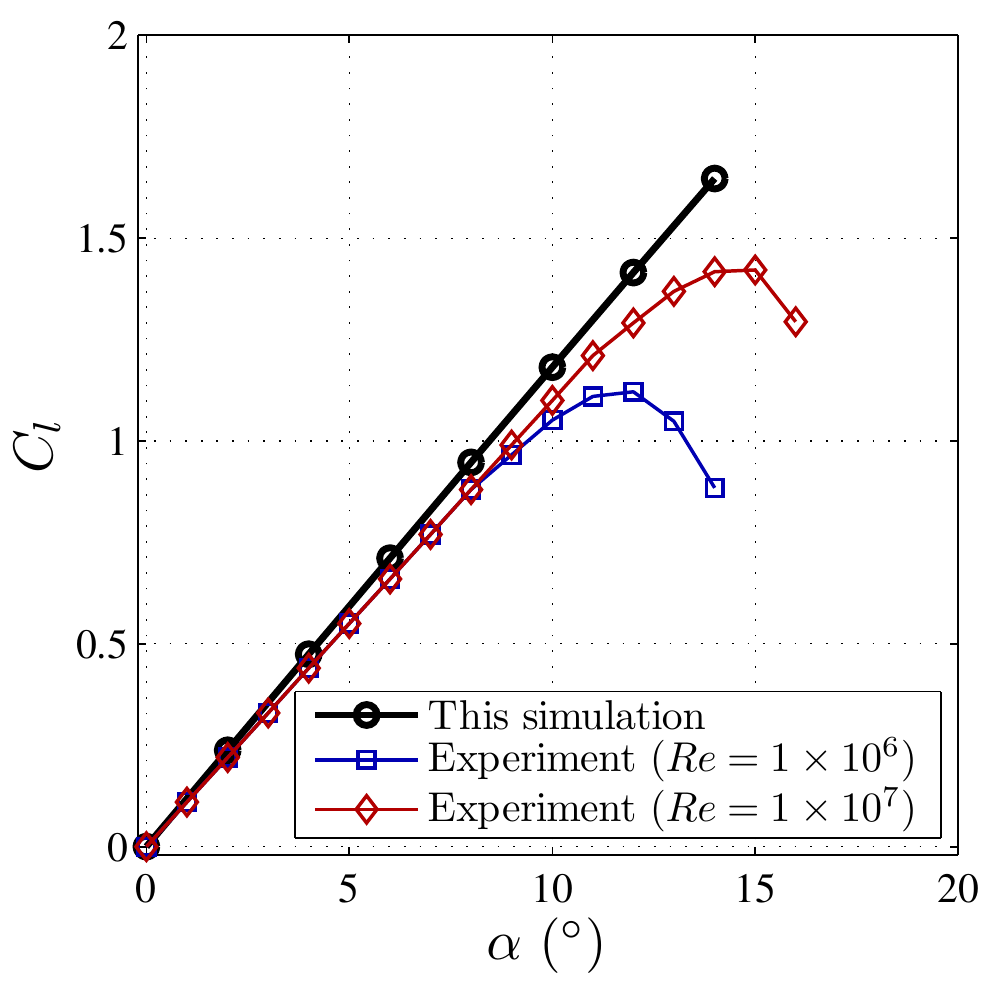}
\end{center} \end{minipage}
\begin{minipage}{0.49\linewidth} \begin{center}  
\includegraphics[width=1\linewidth]{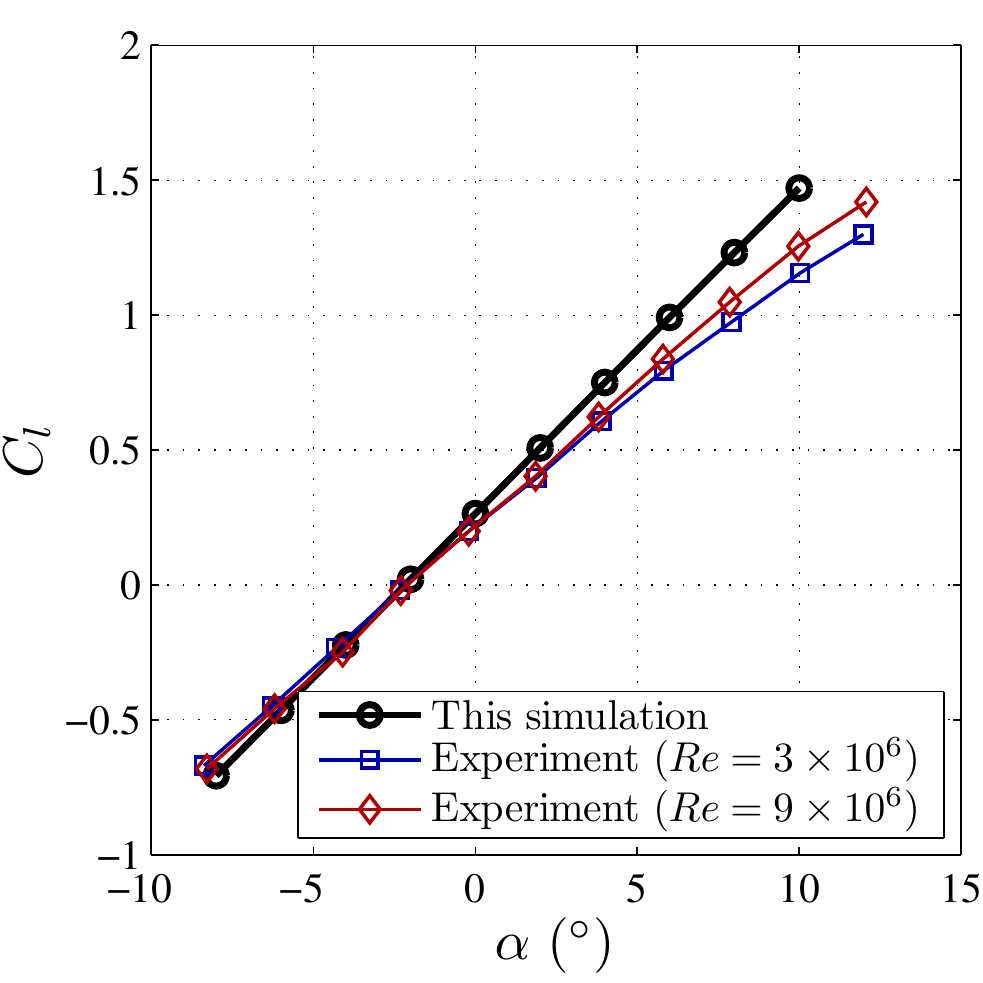}
\end{center} \end{minipage}\\
\vspace{2mm}
\begin{minipage}{0.49\linewidth} \begin{center} (a) NACA 0012 \end{center} \end{minipage}
\begin{minipage}{0.49\linewidth} \begin{center} (b) NACA 2415 \end{center} \end{minipage}

\caption{\small Lift coefficient ($C_l$) vs. angle of attack ($\alpha$) for (a) a NACA 0012 airfoil and (b) a NACA 2415 airfoil. The experimental data for (a) and (b) are from \cite{SheldahlRE:81a} and \cite{AbbottIH:45a}, respectively.} 
\label{fig:lift_steady2}

\end{center}
\end{figure}

\subsection{Airfoils in steady background flow}
 
An impulsively started NACA 0012 airfoil is simulated at various angles of attack. In the body-fixed reference frame, the problem is equivalent to that with a background flow abruptly accelerating from zero to a constant. Although the background flow can be treated as a steady flow for $t>0$, the problem itself is naturally unsteady because of the formation of a starting trailing-edge vortex (TEV). Eventually, a steady flow field around the airfoil can be achieved and the lift will saturate as the starting TEV moves downstream. Therefore, the estimation of the circulation shed from the trailing edge is essential to the accurate prediction of lift generation on the airfoil. For an impulsively started thin airfoil or flat plate, \cite{WagnerH:25a} has provided the numerical data of the time-variant bound circulation, which can be approximated by $\Gamma_b(s^*)/\Gamma_b(\infty) \approx 0.9140-0.3151\mathrm{e}^{-s^*/0.1824}-0.5986\mathrm{e}^{-s^*/2.0282}$, given by \cite{BabinskyH:13a}. $\Gamma_b$ is the total bound circulation; $s^*=s_t/c$ where $s_t$ is the total distance traveled by the airfoil and $c$ is the chord length. Later, \cite{WuZ:15a} modified this function as $\Gamma_b(s^*)/\Gamma_b(\infty) \approx 1-0.8123\mathrm{e}^{-\sqrt{s^*}/1.276}-0.188\mathrm{e}^{-s^*/1.211}+3.2683\times10^{-4}\mathrm{e}^{-s^{*2}/0.892}$ to improve its asymptotic behavior. In this study, the formation of the TEV is solved by implementing the proposed unsteady Kutta condition at each time step. Then, the total bound circulation can be obtained using the Kelvin's circulation theorem as $\Gamma_b = -\Gamma_{TEV}$, where $\Gamma_{TEV}$ is the total circulation of the trailing-edge vortices. $\Gamma_{TEV}$ can be calculated from $\Gamma_{TEV} = \int \dot{\Gamma}_{g} \mathrm{d}t$, with $\dot{\Gamma}_{g}$ determined by equation~(\ref{eq: vorticity_merge8}). Now, we compare the variation of the bound circulation predicted by this model with the approximated Wagner functions in figure~\ref{fig:wagner_comp}. A general good agreement can be observed between this result and the modified Wagner function by \cite{WuZ:15a}, except at early stages ($s^*<5$) where this simulation is slightly different from both Wagner functions. Since Wagner's simulation was based on a flat plate, this difference is likely to reflect the difference of initial vortex shedding between a finite-camber airfoil and a flat plate. In addition, we note that similar to the Wagner function $\Gamma_b(s^*)/\Gamma_b(\infty)$ in this simulation is also independent of the angle of attack, although full data is not presented here for brevity.              

Figure~\ref{fig:lift_steady1}(a) shows the variation of the lift coefficient, $C_l$, for the NACA 0012 airfoil with AoA ranging from 0$^{\circ}$ to 10$^{\circ}$. We can verify the saturation trend of the lift coefficient as $s^*$ increases, which corresponds to the transition of the flow field near the airfoil from unsteady to steady. This transition is also evident in figure~\ref{fig:lift_steady1}(b), which shows the variation of the angle ($\theta_g$) of the trailing-edge vortex sheet. The trend of $\theta_g$ approaching zero also indicates the recovery of condition 4 of the steady-state Kutta condition summarized by \cite{PolingDR:86a} (Section~\ref{sec:6}). Furthermore, it is important to note here that $C_l$ at $s^*=0$ does not start from zero although the bound circulation increases from zero. Following recent studies \citep{Mohseni:13y,WuZ:16a} that attributed major lift generation to the effect of vortex motion, this initial lift should be caused by a strong redistribution effect of the vorticity inside the bound vortex sheet.
          
The steady-state lift coefficients of this study are compared with experimental data for NACA 0012 and NACA 2415 airfoils, as shown in figure~\ref{fig:lift_steady2}. The lift calculations of this model generally match well with experiment at small angles of attack. Furthermore, better agreements can be confirmed for the experimental cases with larger $Re$. This is because larger $Re$ corresponds to smaller mass and momentum deficits associated with the boundary layer, and is therefore better approximated by the vortex-sheet based inviscid flow model. Lastly, the lift stall at larger AoA is not captured because this model does not account for flow separation occurring upstream of the trailing edge.

\begin{figure}
\begin{center}

\begin{minipage}{0.49\linewidth} \begin{center} Experiment \end{center} \end{minipage}
\begin{minipage}{0.49\linewidth} \begin{center} This simulation \end{center} \end{minipage}\\
\vspace{1mm}
\begin{minipage}{0.49\linewidth} 
\begin{center}  
\includegraphics[width=1\linewidth]{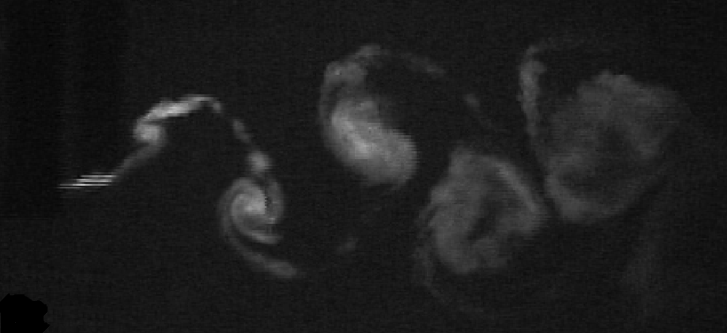}
\end{center} 
\end{minipage}
\begin{minipage}{0.49\linewidth} 
\begin{center}  
\includegraphics[width=0.95\linewidth]{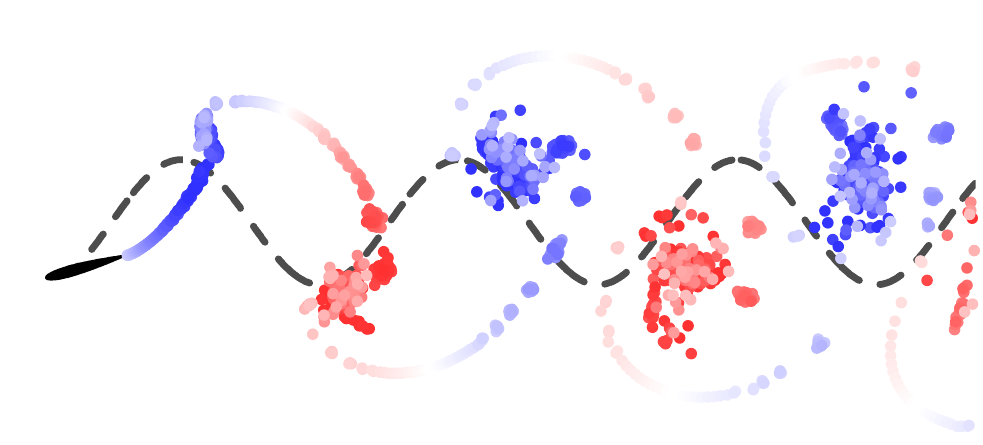}
\end{center}
\end{minipage}

\caption{\small Comparison between flow visualization and simulation for a pitching and heaving NACA 0012 airfoil with $St=0.45$, $\alpha_{max}=30^{\circ}$, and $h_0=0.75c$. The flow visualization image is from \cite{TriantafyllouMS:05a}. The dash line in right plot marks the trajectory of the airfoil.} 
\label{fig:AF_comp1}

\end{center}
\end{figure}
\begin{figure}
\begin{center}

\begin{minipage}{0.49\linewidth} \begin{center} Experiment \end{center} \end{minipage}
\begin{minipage}{0.49\linewidth} \begin{center} This simulation \end{center} \end{minipage}\\
\begin{minipage}{0.49\linewidth} \begin{center}  
\includegraphics[width=.8\linewidth]{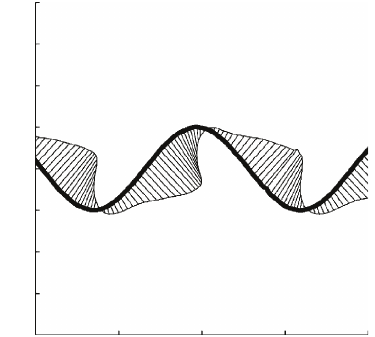}
\end{center} \end{minipage}
\begin{minipage}{0.49\linewidth} \begin{center}  
\vspace{3.5mm} 
\includegraphics[width=.85\linewidth]{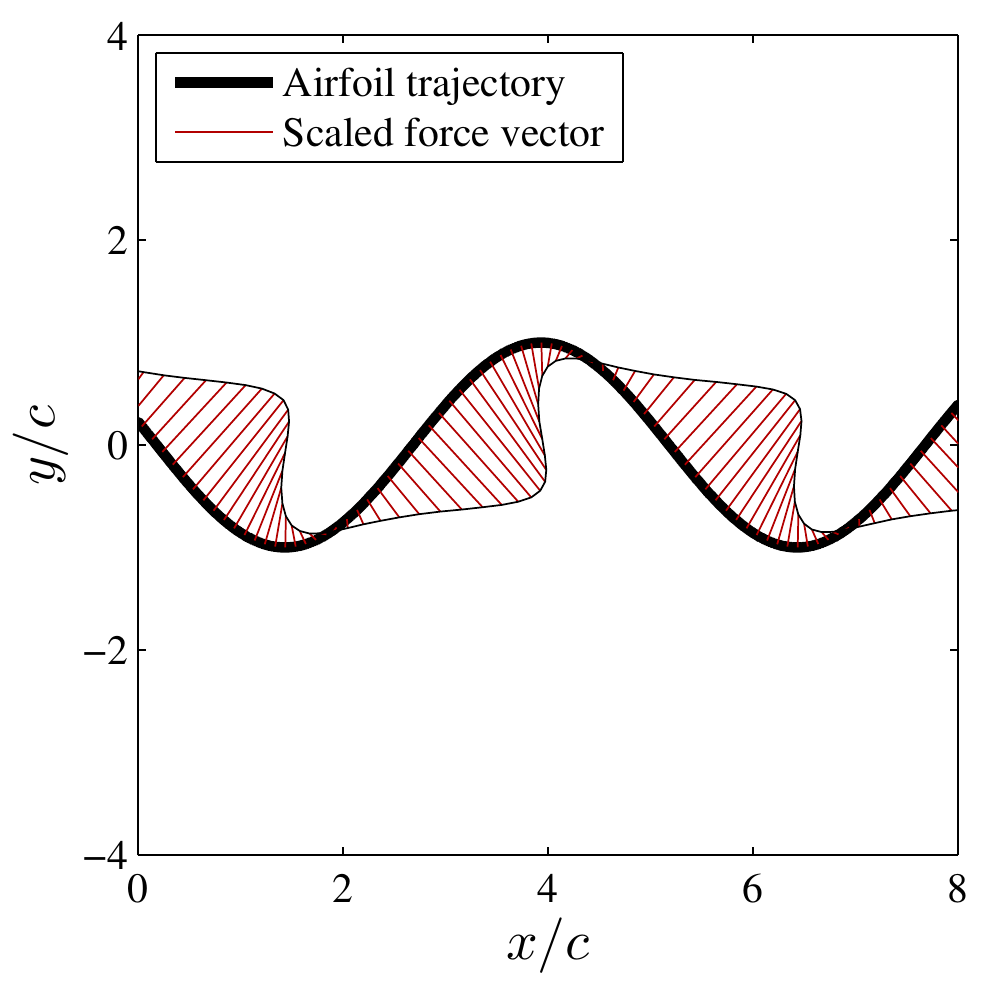}
\end{center} \end{minipage}\\
\begin{minipage}{1\linewidth} \begin{center} (a) $\alpha_{max}=15^{\circ}$ \end{center} \end{minipage}

\begin{minipage}{0.49\linewidth} \begin{center}  
\includegraphics[width=.8\linewidth]{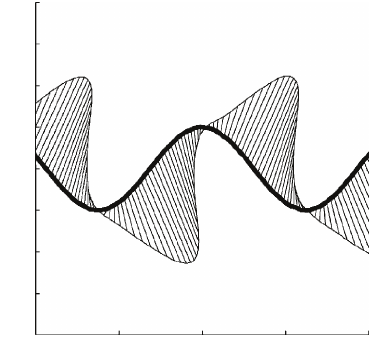}
\end{center} \end{minipage}
\begin{minipage}{0.49\linewidth} \begin{center} 
\vspace{3.5mm} 
\includegraphics[width=.85\linewidth]{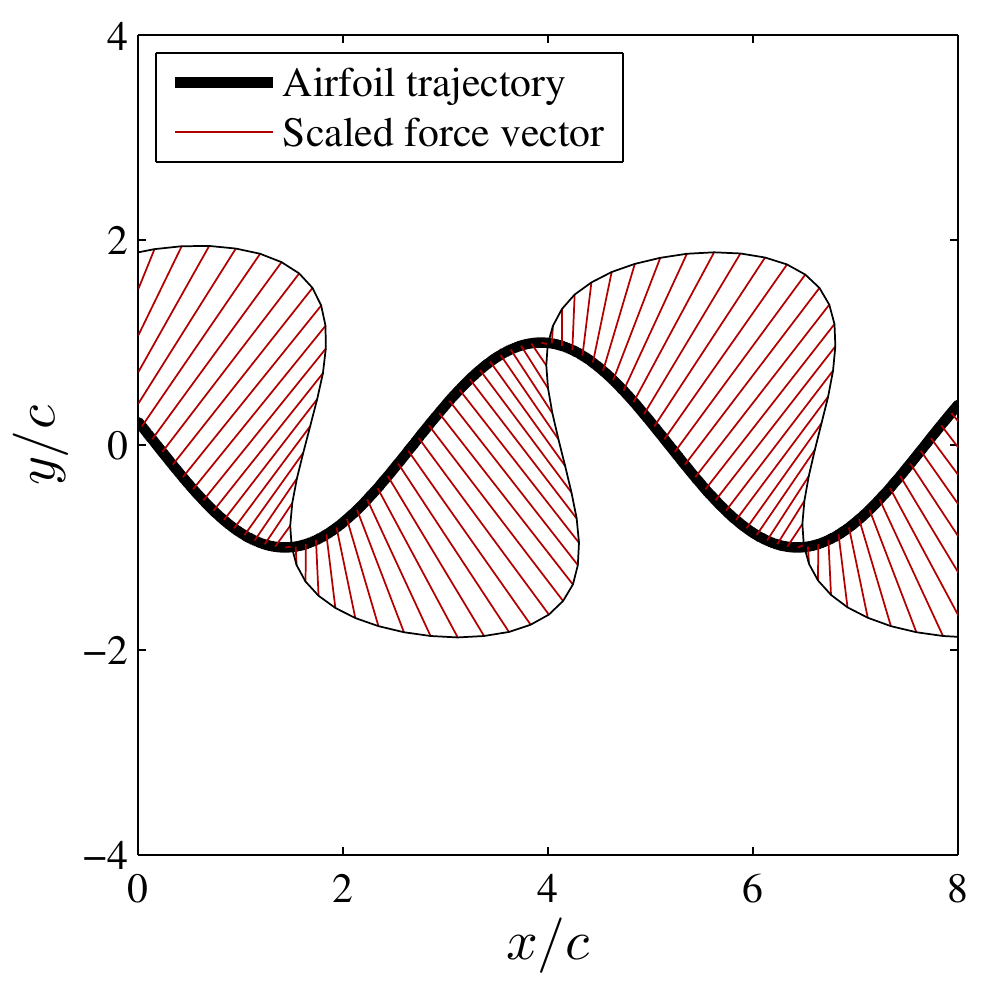}
\end{center} \end{minipage}\\
\begin{minipage}{1\linewidth} \begin{center} (b) $\alpha_{max}=35^{\circ}$ \end{center} \end{minipage}

\caption{\small Comparison of instantaneous thrust vectors between experiment and simulation for a NACA 0012 airfoil. The experimental results are adapted from \cite{TriantafyllouMS:03a}. For both cases, $St=0.4$ and $h_0=c$.} 
\label{fig:force_vector}

\end{center}
\end{figure}
\begin{figure}
\begin{center}

\begin{minipage}{0.49\linewidth} \begin{center} Experiment \end{center} \end{minipage}
\begin{minipage}{0.49\linewidth} \begin{center} This simulation \end{center} \end{minipage}\\

\begin{minipage}{0.49\linewidth} \begin{center}  
\vspace{-10mm} \includegraphics[width=.92\linewidth]{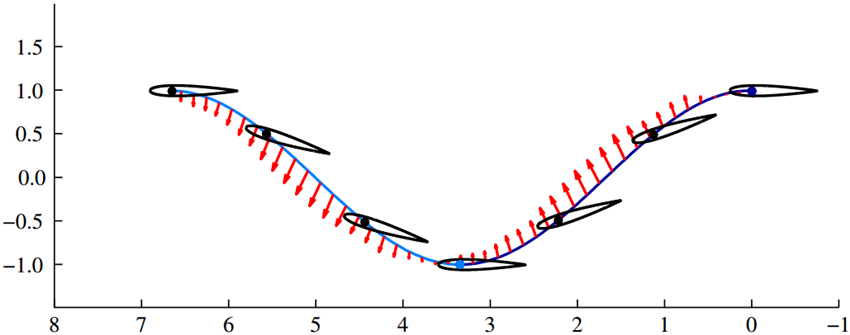}
\end{center} \end{minipage}
\begin{minipage}{0.49\linewidth} \begin{center}  
\includegraphics[width=1.07\linewidth]{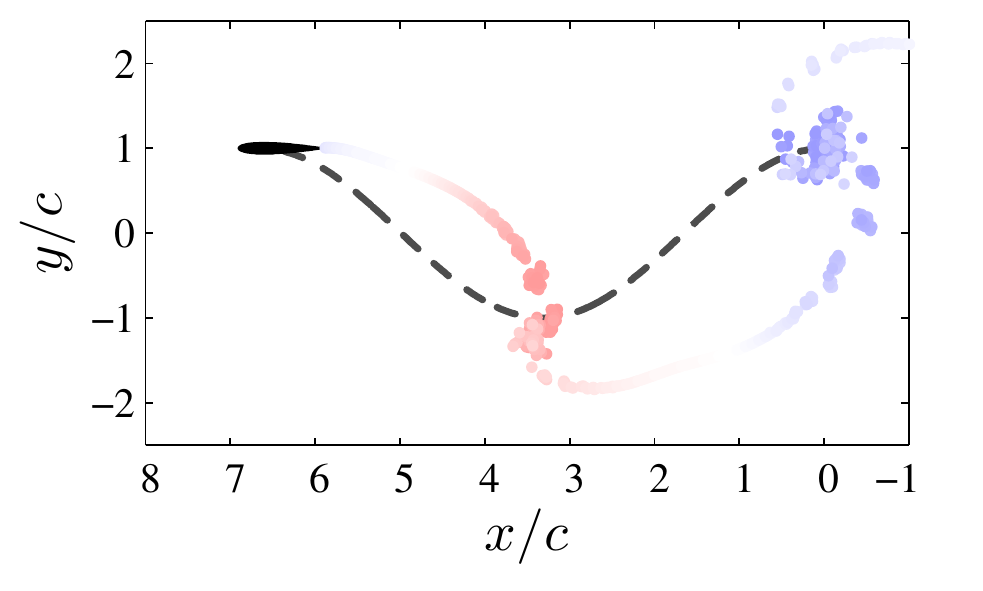}
\end{center} \end{minipage}\\
\begin{minipage}{1\linewidth} \begin{center} (a) Symmetric flapping \end{center} \end{minipage}

\begin{minipage}{0.49\linewidth} \begin{center}  
\vspace{-10mm} \includegraphics[width=.92\linewidth]{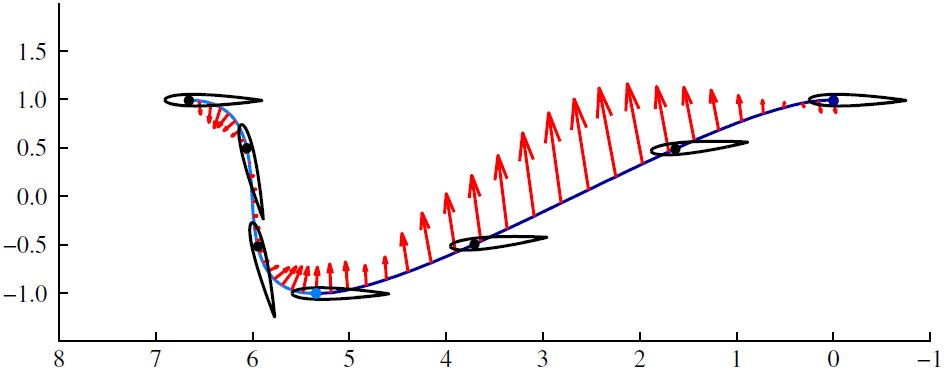}
\end{center} \end{minipage}
\begin{minipage}{0.49\linewidth} \begin{center}  
\includegraphics[width=1.07\linewidth]{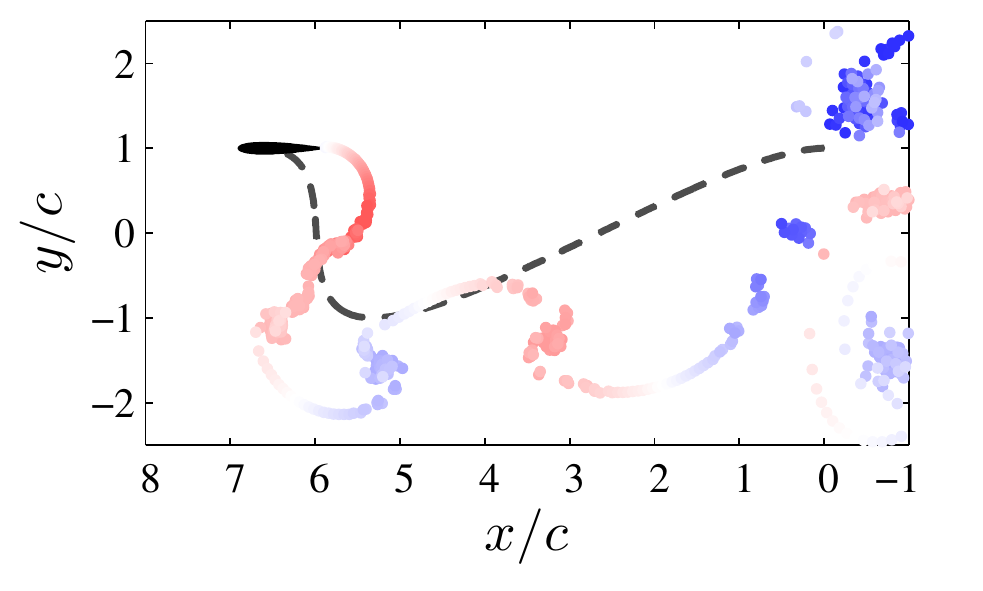}
\end{center} \end{minipage}\\
\begin{minipage}{1\linewidth} \begin{center} (b) Bird-like forward biased downstroke \end{center} \end{minipage}

\begin{minipage}{0.49\linewidth} \begin{center}  
\vspace{-10mm} \includegraphics[width=.92\linewidth]{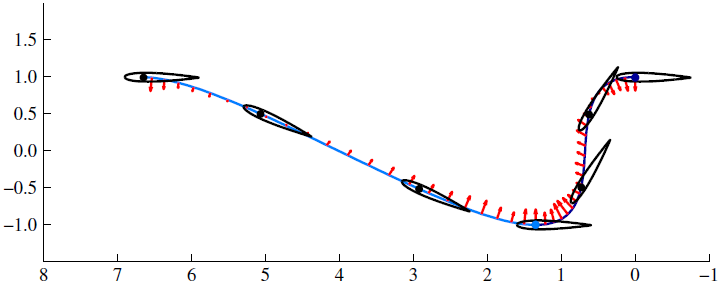}
\end{center} \end{minipage}
\begin{minipage}{0.49\linewidth} \begin{center}  
\includegraphics[width=1.07\linewidth]{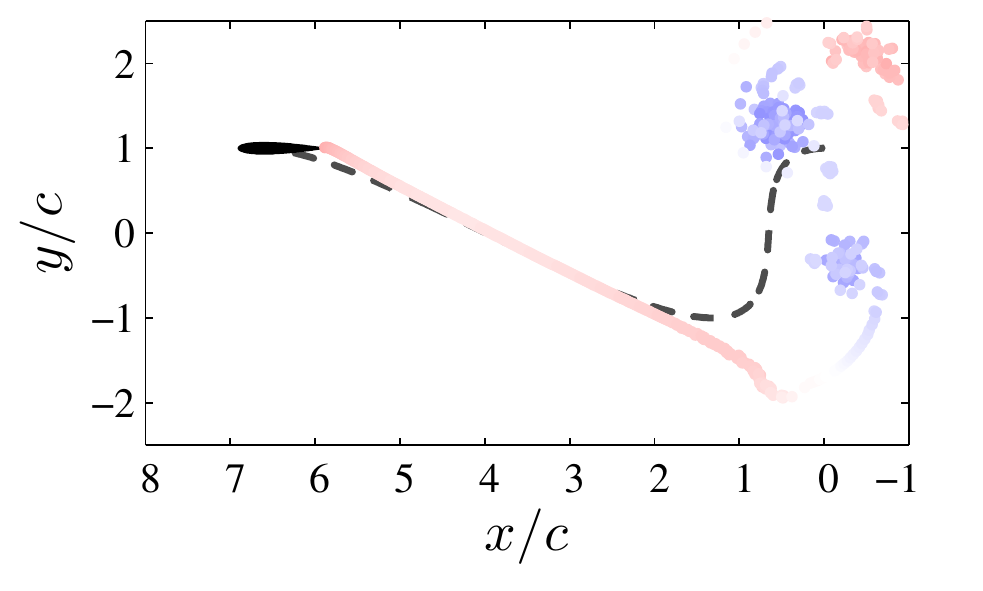}
\end{center} \end{minipage}\\

\begin{minipage}{1\linewidth} \begin{center} (c) Turtle-like backwards moving downstroke \end{center} \end{minipage}

\caption{\small The plots on the left side show the trajectories and force vectors of different unsteady motions of a NACA 0013 airfoil adapted from the experiment in \cite{TriantafyllouMS:14a}. The plots on the right side show the corresponding simulated wake patterns.} 
\label{fig:AF_comp2}

\end{center}
\end{figure}
\begin{figure}
\begin{center}

\begin{minipage}{0.49\linewidth} \begin{center}  
\includegraphics[width=1\linewidth]{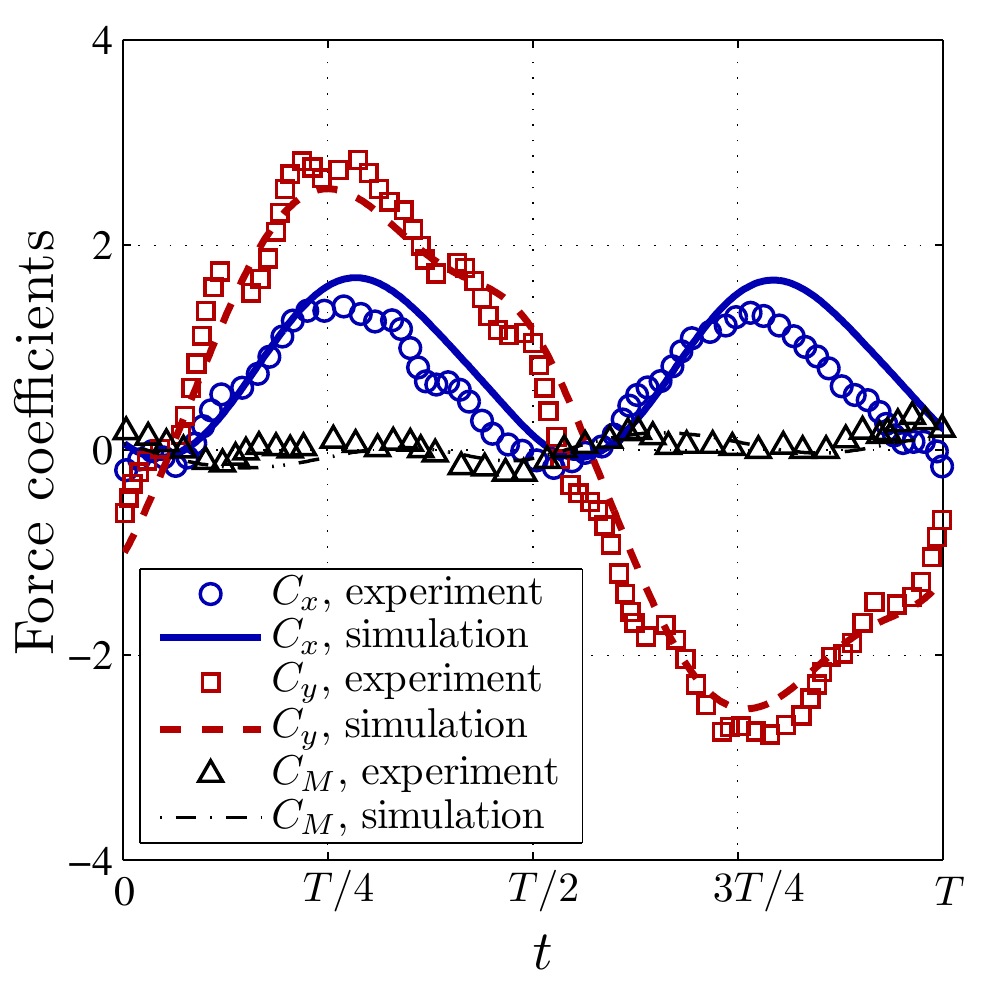}
\end{center} \end{minipage}
\begin{minipage}{0.49\linewidth} \begin{center}  
\includegraphics[width=1\linewidth]{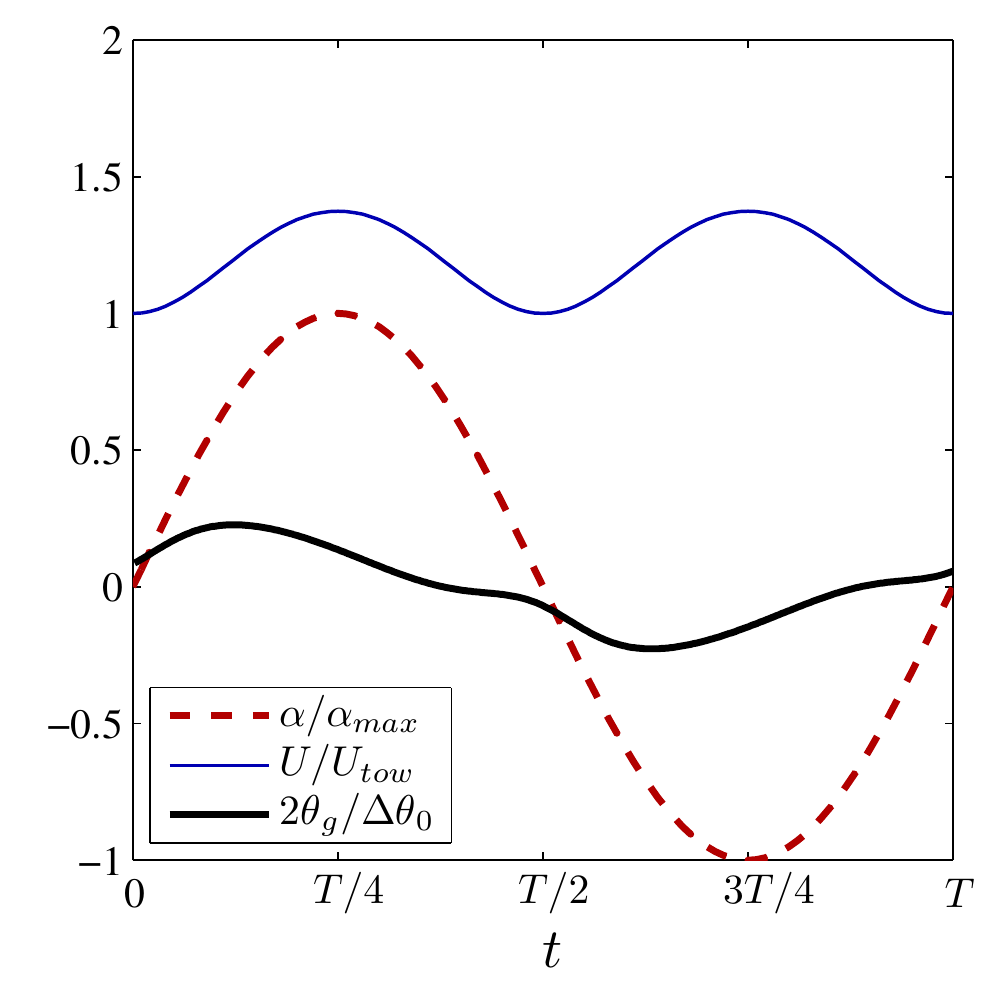}
\end{center} \end{minipage}\\

\begin{minipage}{0.49\linewidth} \begin{center} (a) \end{center} \end{minipage}
\begin{minipage}{0.49\linewidth} \begin{center} (b) \end{center} \end{minipage}

\caption{\small Result of the symmetric flapping motion corresponding to figure~\ref{fig:AF_comp2}(a). (a) Comparison between the measured force coefficients of \cite{TriantafyllouMS:14a} and the estimated force coefficients from this simulation. (b) Variations of $\alpha$, $U$, and $\theta_g$ during one cycle.} 
\label{fig:AF_comp3}

\end{center}
\end{figure}
\begin{figure}
\begin{center}

\begin{minipage}{0.49\linewidth} \begin{center}  
\includegraphics[width=1\linewidth]{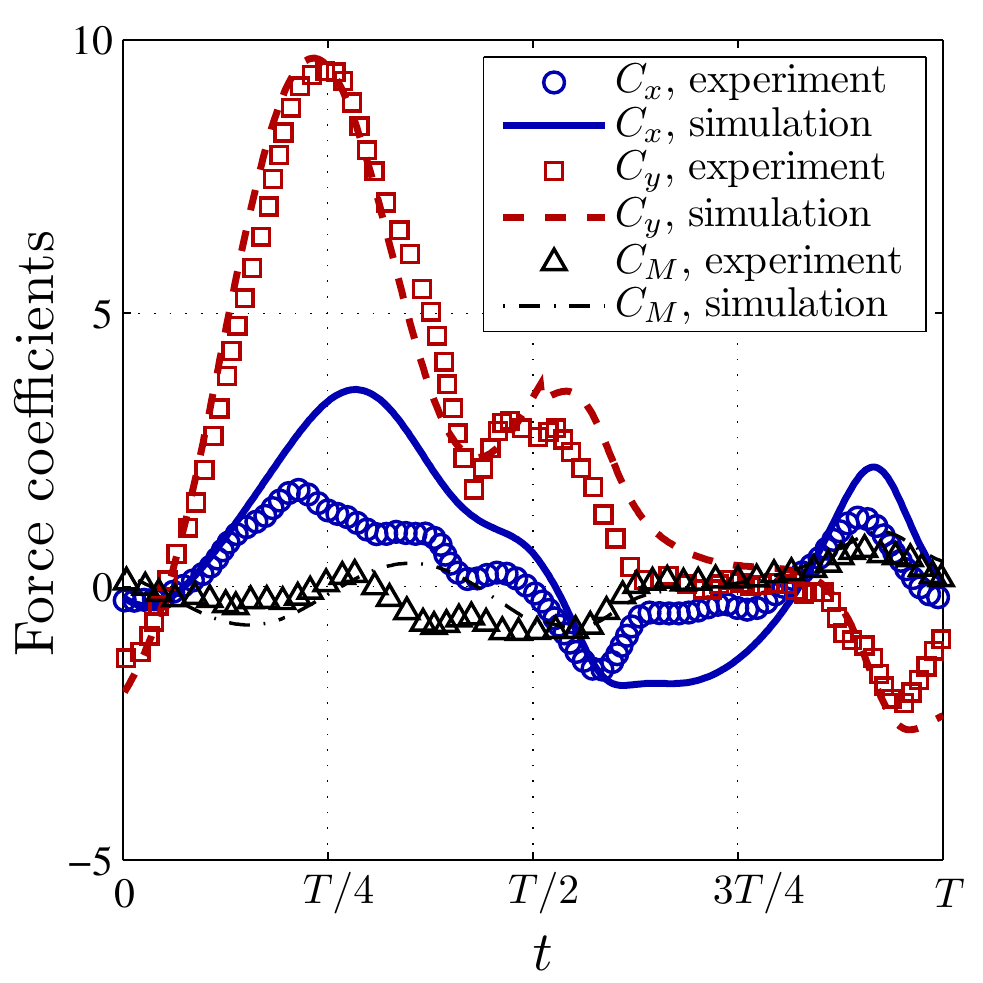}
\end{center} \end{minipage}
\begin{minipage}{0.49\linewidth} \begin{center}  
\includegraphics[width=1\linewidth]{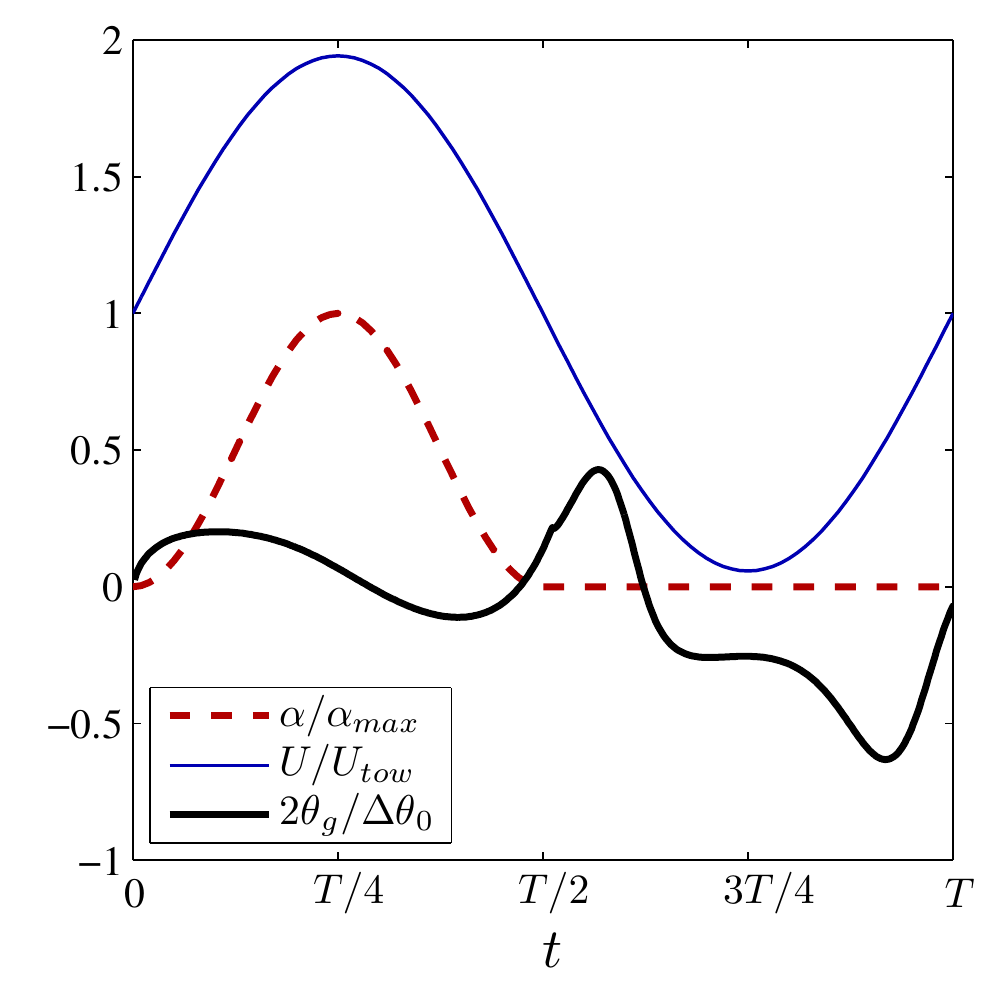}
\end{center} \end{minipage}\\

\begin{minipage}{0.49\linewidth} \begin{center} (a) \end{center} \end{minipage}
\begin{minipage}{0.49\linewidth} \begin{center} (b) \end{center} \end{minipage}

\caption{\small Result of the bird-like forward biased downstroke corresponding to figure~\ref{fig:AF_comp2}(b). (a) Comparison between measured and estimated force coefficients. (b) Variations of $\alpha$, $U$, and $\theta_g$ during one cycle.} 
\label{fig:AF_comp4}

\end{center}
\end{figure}
\begin{figure}
\begin{center}

\begin{minipage}{0.49\linewidth} \begin{center}  
\includegraphics[width=1\linewidth]{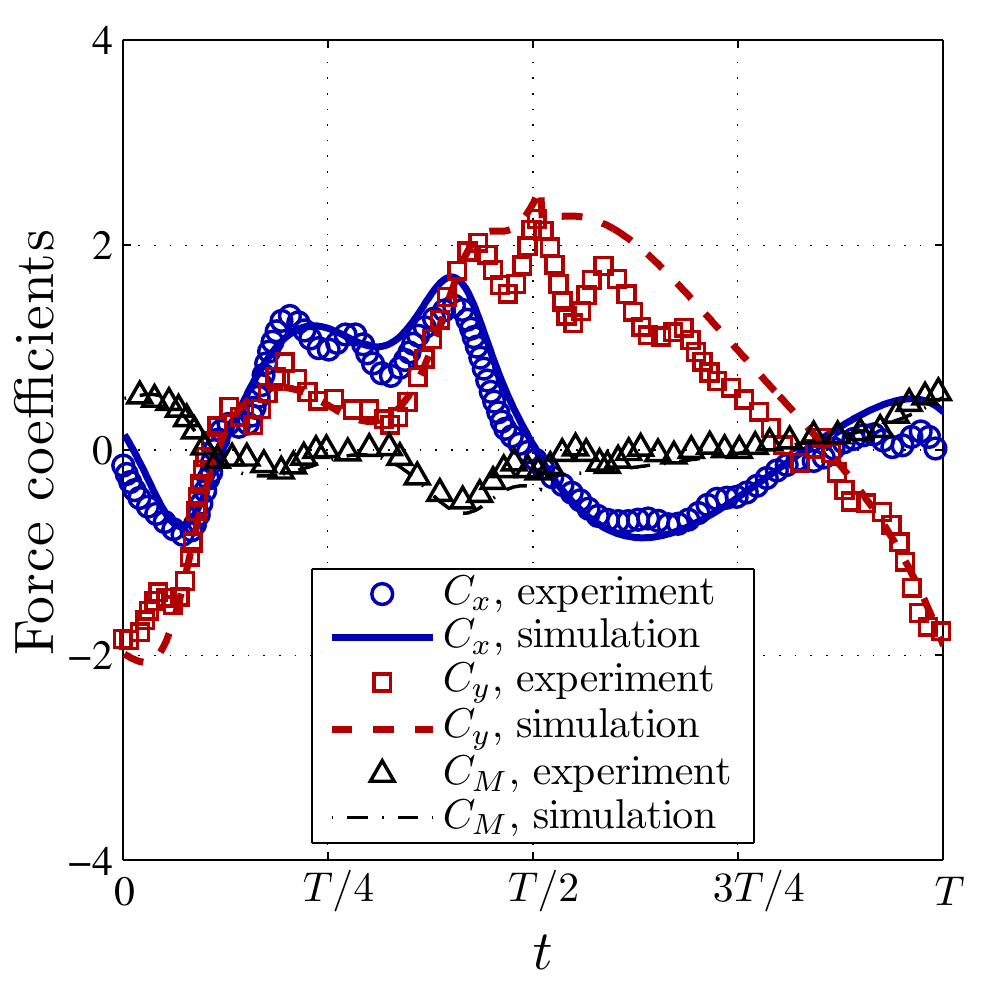}
\end{center} \end{minipage}
\begin{minipage}{0.49\linewidth} \begin{center}  
\includegraphics[width=1\linewidth]{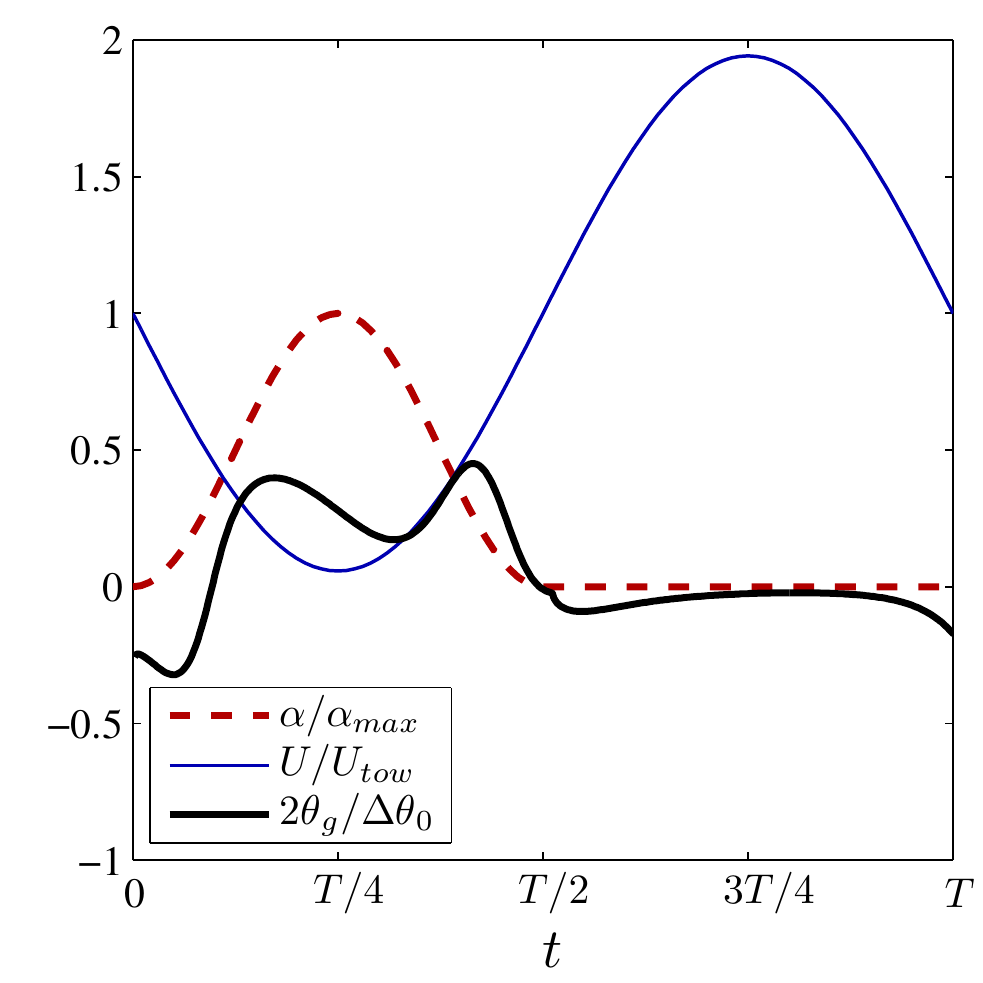}
\end{center} \end{minipage}\\

\begin{minipage}{0.49\linewidth} \begin{center} (a) \end{center} \end{minipage}
\begin{minipage}{0.49\linewidth} \begin{center} (b) \end{center} \end{minipage}

\caption{\small Result of the turtle-like backwards moving downstroke corresponding to figure~\ref{fig:AF_comp2}(c). (a) Comparison between measured and estimated force coefficients. (b) Variations of $\alpha$, $U$, and $\theta_g$ during one cycle.} 
\label{fig:AF_comp5}

\end{center}
\end{figure}

\subsection{Airfoils with unsteady motions}

Next, the performance of this vortex-sheet based aerodynamic model is further justified by simulating a series of unsteady motions of the NACA airfoils. We first investigate a NACA 0012 airfoil with a combined pitching and heaving motion adapted from the experiment of \cite{TriantafyllouMS:03a}. For all tests, the chord length and the towing speed are fixed at $c=0.1$ m and $U_{tow}=0.4$ m/s, respectively. The corresponding Reynolds number is $4\times10^4$. The pivot for the pitching motion is fixed at 1/3 chord. The phase difference angle between the pitching and heaving motions is set to 90$^\circ$. The characteristic parameters for this motion are the Strouhal number, $St$, the amplitude of angle-of-attack, $\alpha_{max}$, and the heave amplitude, $h_0$, which could be adjusted by controlling the pitching and heaving motions. Figure~\ref{fig:AF_comp1} compares the wake structures between this simulation and the flow visualization for a sample case ($St=0.45$, $\alpha_{max}=30^{\circ}$, and $h_0=0.75c$). The matching of the wake patterns between experiment and simulation is promising. Figure~\ref{fig:force_vector} further plots the instantaneous force vectors along the trajectories of two different pitching and heaving motions. The results demonstrate reasonable agreement of the force magnitude and direction between experiment and simulation. This quantitatively validates the performance of the aerodynamic model and the TEV formation conditions for airfoils undergoing unsteady motions. However, since the LEV shedding has not been considered here, the simulations with larger $\alpha_{max}$ or $St$ values tend to overestimate the force due to possible flow separation after the leading edge.

The unsteadiness of the airfoil motion can be further increased by adding an oscillatory in-line motion on top of the pitching and heaving motion introduced above. Two typical such motions were experimentally studied by \cite{TriantafyllouMS:14a}, namely, the bird-like forward biased downstroke and the turtle-like backwards moving downstroke. The trajectories of these two motions are shown in figure~\ref{fig:AF_comp2}(b) and (c), with the simulated flow field showing the vortical structures in the wake. For comparison, a symmetric flapping case without any additional in-line motion is shown in figure~\ref{fig:AF_comp2}(a). The airfoil investigated here is a NACA 0013 type with $c=0.055$ m and the pivot at the quarter-chord. The Reynolds number is fixed at 11000 which corresponds to a constant towing speed of $U_{tow}=0.2$ m/s. For the pitching and heaving motions of all cases, the characteristic parameters are $St=0.3$, $\alpha_{max}=25^{\circ}$, and $h_0 =c$. The controlling parameter here is the stroke angle, $\beta$, associated with the added in-line motion. $\beta$ is defined based on the $x$ and $y$ positions of the airfoil in the carriage reference frame. The interested readers are referred to \cite{TriantafyllouMS:14a} for more details of the original experiment.

For quantitative comparison, the force and torque coefficients are estimated for the unsteady motions presented in figure~\ref{fig:AF_comp2}. Similar to \cite{TriantafyllouMS:14a}, the force coefficients in the $x$ and $y$ directions together with the torque coefficient are defined as $C_x=2F_x(\rho U_{tow}^2c)^{-1}$, $C_y=2F_y(\rho U_{tow}^2c)^{-1}$, and $C_M=2T_{\tau}(\rho U_{tow}^2c^2)^{-1}$, respectively, where $F_x$, $F_y$, and $T_{\tau}$ are computed from equations~(\ref{eq:AF_force}) and~(\ref{eq:AF_torque}). The evolution of $C_x$, $C_y$, and $C_M$ during each cycle of the prescribed unsteady motions are compared with the experiment data in figures~\ref{fig:AF_comp3}(a),~\ref{fig:AF_comp4}(a), and~\ref{fig:AF_comp5}(a). We again observe a generally good agreement between experiment and simulation, verifying the performance of the proposed flow model. Especially, the results of $C_y$ have promising accuracy for all three different cases. Since $C_y$ physically represents the lift coefficient, this indicates a prospective application of the current model for lift estimation without modeling the leading-edge separation. However, the void of flow separation in the current simulation seems to have a notable impact on $C_x$, which corresponds to the thrust coefficient. This is reflected by the over-prediction of $C_x$ in some cases displayed in figures~\ref{fig:AF_comp3}(a) and~\ref{fig:AF_comp4}(a). Other than the flow separation, the viscous effect at the solid-fluid interface could also affect the accuracy of predicting thrust or drag using an inviscid flow model. 

To explain the effect of the additional in-line motion on force generation of the airfoil, the variations of the airfoil velocity $U$ and the angle of attack $\alpha$ are plotted in figures~\ref{fig:AF_comp3}(b),~\ref{fig:AF_comp4}(b), and~\ref{fig:AF_comp5}(b). We can observe that the variations of $\alpha$ in figures~\ref{fig:AF_comp4} and~\ref{fig:AF_comp5} are identical, and $\alpha$ only changes in the first half cycle while it remains zero in the second half cycle. Since the shedding of strong vorticity mainly occurs at non-zero angles of attack, the force generation associated with vortex shedding should mostly happen during the first half cycle. In this sense, the first half cycle is the actual `stroke' while the second half can be considered as the `recovery'. However, the different in-line motions during the first half cycle causes $U$ to increase significantly for the bird-like downstroke in figure~\ref{fig:AF_comp4} and decrease significantly for the turtle-like downstroke in figure~\ref{fig:AF_comp5}. This creates stronger and faster trailing-edge vortices of the bird-like downstroke compared to the turtle-like downstroke. As a result, the bird-like downstroke provides much higher lift than the turtle-like downstroke. Finally, we note that the angle of the trailing-edge vortex sheet, $\theta_g$, varies smoothly within the limits of the two tangential directions of the trailing edge, $-\Delta \theta_0/2$ and $\Delta \theta_0/2$. This also implies the correct implementation of the proposed models, and is in accordance with the experimental observation of \cite{PolingDR:86a} that the direction of the trailing-edge streamline changes smoothly.     
\section{Conclusions}\label{sec:8}

An unsteady aerodynamic model for an airfoil was derived based on the dynamics of the bound vortex sheet and the wake vortices. The vorticity generation mechanism at the trailing edge was studied since it is essential to predict the vortex shedding and evolution processes in the wake. For a flat plate or a cusped trailing edge, this can be solved by applying an unsteady Kutta condition, based on the physical sense that flow cannot turn around a sharp edge. However, this condition for an airfoil with finite-angle trailing edge is not straightforward to implement. Specifically, the vortex sheet formed at the trailing-edge of a flat plate is known to the tangential to the flat plate, whereas the angle of the forming vortex sheet for an airfoil could vary between the two tangential directions of the trailing edge. Realizing that any arbitrary choice for the vortex-sheet angle would be ad-hoc, this study proposed to calculate this angle based on the basic conservation laws of mass and momentum, together with the condition derived from the Kelvin's circulation theorem. This resulted in the analytical expression of the angle, strength, and velocity of a free vortex sheet formed at a finite-angle trailing edge, establishing a general unsteady Kutta condition for relevant problems. The significance of this work is that the vortex-sheet formation condition allows the angle of the forming vortex sheet to continuously change between the two tangents of the trailing edge. This resolves the paradox of the Giesing-Maskell model that it does not converge to the steady-state Kutta condition. Airfoils in various steady and unsteady flows were simulated and the resulting flow field and force calculations were compared with experimental data. The promising agreement between simulation and experiment confirmed the validity of the proposed unsteady Kutta condition as well as the vortex-sheet based aerodynamic model. 

\section{Acknowledgments}
\begin{acknowledgments}
This work is supported by a grant from the Office of Naval Research. We would also like to thank Dr. Adam DeVoria for providing helpful discussions and comments.
\end{acknowledgments}
\appendix
\section{} \label{sec:9}
\begin{figure}
\begin{center}
\scalebox{0.7}{\includegraphics{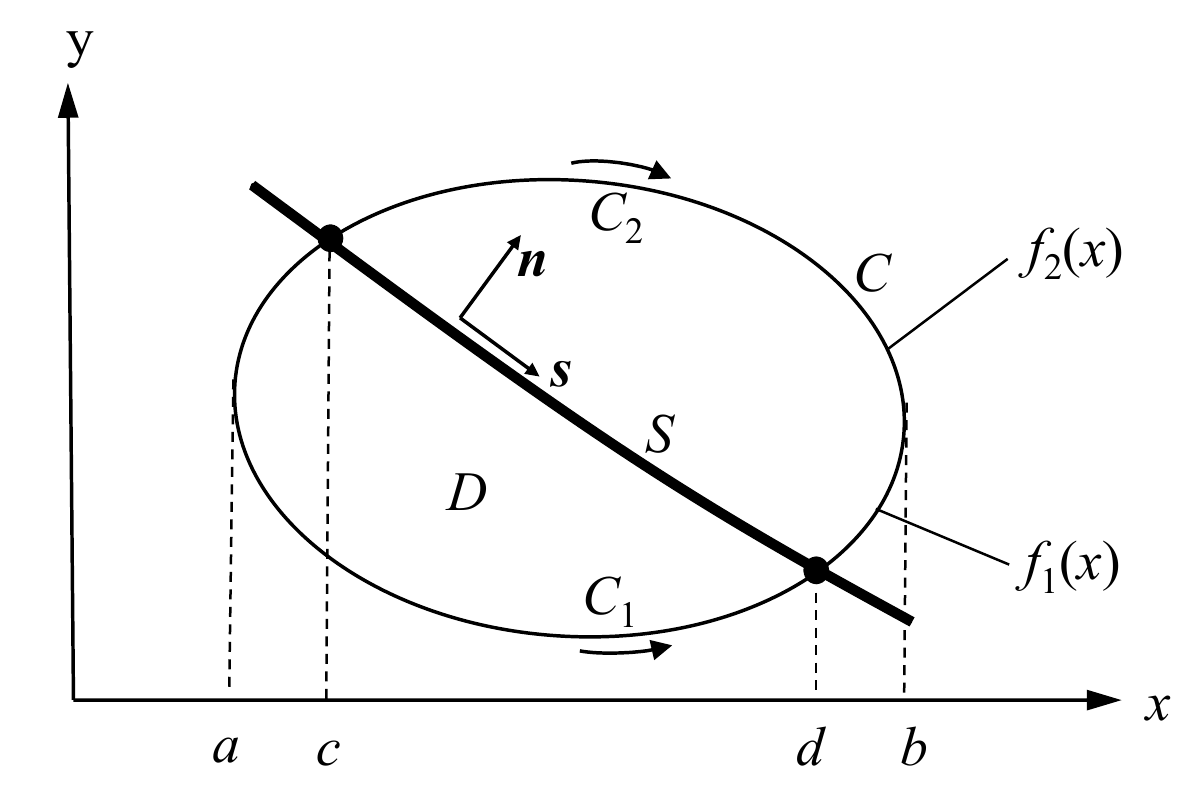}}
\caption{Domain $D$ divided by a discontinuous interface $S$.}
\label{fig:stokes}
\end{center}
\end{figure}
This appendix derives the Green's theorem used in equation~(\ref{eq: tot_circulation}) for a domain containing discontinuous interfaces. The process is similar to the proof of the original Green's theorem \citep{KaplanW:02a}. We start with the simple case where a 2D domian $D$ is enclosed by a smooth simple closed curve $C$, as shown in figure~\ref{fig:stokes}. $P(x,y)$ and $Q(x,y)$ are continuous functions and have continuous first partial derivatives in $D$, except on a dividing interface $S$ ($S = \{(x,f_s(x)), c \leq x \leq d\}$). $C$ can be divided into $C_1$ ($C_1 = \{(x,f_1(x)), a \leq x \leq b\}$) and $C_2$ ($C_2 = \{(x,f_2(x)), a \leq x \leq b\}$).  

Now, the first integral to evaluate is 
\begin{equation}
\label{eq: APX0_Int1}
\begin{split}
 \iint_D \frac{\partial P}{\partial y} \mathrm{d}x\mathrm{d}y & = \int_{a}^{b}\int_{f_1(x)}^{f_2(x)} \frac{\partial P}{\partial y} \mathrm{d}y\mathrm{d}x\\
 & = \int_{a}^{c} [P(x,f_2(x))-P(x,f_1(x))] \mathrm{d}x + \int_{d}^{b} [P(x,f_2(x))-P(x,f_1(x))] \mathrm{d}x\\
 & + \int_{c}^{d} \left[P(x,f_2(x))-P(x,f_s^+(x)) + \int_{f_s^-(x)}^{f_s^+(x)} \frac{\partial P}{\partial y} \mathrm{d}y + P(x,f_s^-(x))-P(x,f_1(x))\right] \mathrm{d}x\\
 & = \int_{a}^{b} [P(x,f_2(x))-P(x,f_1(x))] \mathrm{d}x + \int_{c}^{d} \left[ -[\![ P(x,f_s(x)) ]\!] + \int_{f_s^-(x)}^{f_s^+(x)} \frac{\partial P}{\partial y} \mathrm{d}y \right] \mathrm{d}x,
\end{split}
\end{equation} 
where $f_s^+(x)$ and $f_s^-(x)$ represents the upper and lower limits of $f_s(x)$, and the jump term $[\![ P(x,f_s(x)) ]\!] = P(x,f_s^+(x))-P(x,f_s^-(x))$. Note here, $\partial P/\partial y$ is not well defined on $S$ and is dependent on the physical problem. In general, we assume $\partial P/\partial y$ to be finite on $S$ so equation~(\ref{eq: APX0_Int1}) takes the form
\begin{equation}
\label{eq: APX0_Int2}
 \iint_D \frac{\partial P}{\partial y} \mathrm{d}x\mathrm{d}y = - \oint_C P \mathrm{d}x - \int_S [\![ P ]\!] \mathrm{d}x.   
\end{equation} 
Similarly, $\iint_D \frac{\partial Q}{\partial x} \mathrm{d}x\mathrm{d}y$ can be derived as
\begin{equation}
\label{eq: APX0_Int3}
 \iint_D \frac{\partial Q}{\partial x} \mathrm{d}x\mathrm{d}y = \oint_C Q \mathrm{d}y + \int_S [\![ Q ]\!] \mathrm{d}y.   
\end{equation} 
Therefore, combining equations~(\ref{eq: APX0_Int2}) and~(\ref{eq: APX0_Int3}) yields a general Green's theorem for a domain with a discontinuous interface:
\begin{equation}
\label{eq: APX0_Int4}
 \iint_D \left( \frac{\partial Q}{\partial x} - \frac{\partial P}{\partial y} \right)\mathrm{d}x\mathrm{d}y = \oint_C (P \mathrm{d}x + Q \mathrm{d}y) + \int_S ([\![ P ]\!] \mathrm{d}x + [\![ Q ]\!] \mathrm{d}y).   
\end{equation}
Apparently, the discontinuity associated with the interface $S$ causes an additional jump term on the right hand side of the original Green's theorem. 

However, here we consider a special case where the derivatives of $P$ and $Q$ on $S$ are not finite and have the form
\begin{equation}
\label{eq: APX0_Int5}
 \begin{dcases} 
 \frac{\partial P}{\partial n} &= [\![P(s)]\!] \delta(n)\\
 \frac{\partial Q}{\partial n} &= [\![Q(s)]\!] \delta(n). 
\end{dcases}  
\end{equation}
where $\delta$ is the Dirac delta function, $n$ and $s$ are the normal and tangential coordinates of $S$, respectively. Therefore, $\partial P/\partial y$ on $S$ can be written as
\begin{equation}
\label{eq: APX0_Int6}
 \frac{\partial P}{\partial y} = [\![ P(x,f_s(x)) ]\!] \delta(y-f_s(x)) \hspace{5mm} \text{for} \hspace{5mm} (x,y) \in S.	 
\end{equation} 
Now, equation~(\ref{eq: APX0_Int6}) can be plugged into equation~(\ref{eq: APX0_Int1}) to obtain
\begin{equation}
\label{eq: APX0_Int7}
 \iint_D \frac{\partial P}{\partial y} \mathrm{d}x\mathrm{d}y = - \oint_C P \mathrm{d}x.   
\end{equation} 
In the same way, the counterpart for equation~(\ref{eq: APX0_Int3}) becomes
\begin{equation}
\label{eq: APX0_Int8}
 \iint_D \frac{\partial Q}{\partial x} \mathrm{d}x\mathrm{d}y = \oint_C Q \mathrm{d}y.	 
\end{equation} 
Thus, the Green's theorem in this case has its original form
\begin{equation}
\label{eq: APX0_Int9}
 \iint_D \left( \frac{\partial Q}{\partial x} - \frac{\partial P}{\partial y} \right)\mathrm{d}x\mathrm{d}y = \oint_C (P \mathrm{d}x + Q \mathrm{d}y).   
\end{equation}
Finally, applying domain decomposition similar to that in \cite{KaplanW:02a}, the above results can be extended to a general-shaped volume with multiple discontinuous surfaces. 

\section{} \label{sec:10}

\begin{figure}
\begin{center}
\scalebox{0.7}{\includegraphics{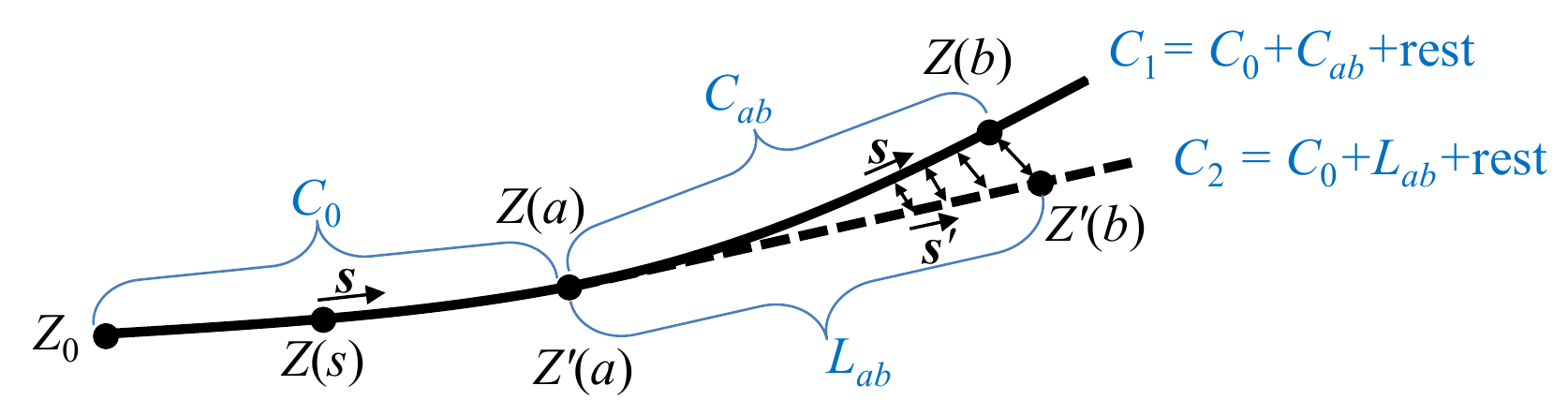}}
\caption{A diagram of the curves $C_1$ and $C_2$ for calculating the integral $\int \gamma(s)(Z_0-Z(s))^{-1} \mathrm{d}s$.}
\label{fig:TE_AF_JFM1}
\end{center}
\end{figure}

This appendix, together with appendix~\ref{sec:11}, provides the detailed calculation for equation~(\ref{eq: VS_vel_TE3}). In this study, we are faced with the task of computing the line integral $\int \gamma(s)(Z_0-Z(s))^{-1} \mathrm{d}s$ between two different points, $Z(a)$ and $Z(b)$ ($0<a<b$), along a simple open (without self-intersection) curve, $C_1$, which starts from $Z_0$ in the complex domain, as shown in figure~\ref{fig:TE_AF_JFM1}. $s$ is the curve length between $Z_0$ and an arbitrary point ($Z(s) = X(s) + iY(s)$) on $C_1$, so $Z_0 = Z(0)$. $\gamma(s)$ is a real function defined on curve $C_1$. Given $C_1$ and $\gamma(s)$ are smooth (class $C^{\infty}$) for $a \leq s \leq b$, it means that $X(s)$, $Y(s)$, and $\gamma(s)$ are infinitely differentiable on $C_{ab} = \{Z(s), a \leq s \leq b\}$. 

We start by constructing a straight line $L$ which begins from $Z(a)$ and shares the tangential direction of $C_1$ at $Z(a)$. Therefore, $L$ is mathematically given by the following conditions: $Z'(a) = Z(a)$ and $(\mathrm{d}Z'(s')/\mathrm{d}s')|_{s'=a} = (\mathrm{d}Z(s)/\mathrm{d}s)|_{s=a}$, where $Z'(s') = X'(s') + iY'(s')$ denotes an arbitrary point on curve $C_2 = C_0 + L$ ($C_0 = \{Z(s), 0 \leq s < a\}$) and $s'$ is the curve length between $Z_0$ and $Z'(s')$. Now, a one-to-one mapping can be readily established between $C_1$ and $C_2$ by setting $s=s'$. In this way, $\gamma$ for $C_1$ can be mapped to $\gamma'$ for $C_2$ with the relation $\gamma'(s') = \gamma(s')$, so the original integral becomes
\begin{equation}
\label{eq: APX_Int0}
 \int_{C_{ab}} \frac{\gamma(s) \mathrm{d}s}{Z_0-Z(s)} = \int_{L_{ab}} \frac{\gamma'(s') \mathrm{d}s'}{Z_0-Z(s')},
\end{equation}     
where $L_{ab} = \{Z'(s'), a \leq s' \leq b\}$. This allows us to change the integration path from a curve between $Z(a)$ and $Z(b)$ to a straight line between $Z'(a)$ and $Z'(b)$. Noting that $\gamma'(s') = \gamma(s')$, we can further simplify the result of equation~(\ref{eq: APX_Int0}) and divide it into
\begin{equation}
\label{eq: APX_Int1}
 \int_{L_{ab}} \frac{\gamma(a) \mathrm{d}s'}{Z_0-Z'(s')} + \int_{L_{ab}} \frac{(\gamma(s')-\gamma(a)) \mathrm{d}s'}{Z_0-Z'(s')} + \int_{L_{ab}} \left( \frac{1}{Z_0-Z(s')}-\frac{1}{Z_0-Z'(s')} \right) \gamma(s') \mathrm{d}s',
\end{equation}
where the three terms will be treated separately in the following part.  

To calculate the first term in equation~(\ref{eq: APX_Int1}), we set $z' = Z'(s')$ for $z' \in L_{ab}$, so $\mathrm{d}s' = e^{-i\theta_a} \mathrm{d}z'$ where $\theta_a$ is the angle of the tangential direction of $C_1$ or $C_2$ at $Z(a)$. As a result, this term can be integrated as
\begin{equation}
\label{eq: APX_Int2}
 \int_{L_{ab}} \frac{\gamma(a) \mathrm{d}s'}{Z_0-Z'(s')} = \int_{L_{ab}} \frac{\gamma(a) e^{-i\theta_a} \mathrm{d}z'}{Z_0-z'} = - \gamma(a) e^{-i\theta_a} \left.\left[\vphantom{\int} \ln(z'-Z_0) \right] \right|_{Z(a)}^{Z'(b)}.
\end{equation}
So far, the derivations have been performed for two arbitrary points $Z_a$ and $Z_b$ ($0<a<b$) on $C_1$. In the problem of interest, we are concerned with the limiting results when $Z(a) \rightarrow Z_0$ or $a \rightarrow 0$. In this case, together with the relation $z'-Z(a) = e^{i\theta_a} (s'-a)$, the result of equation~(\ref{eq: APX_Int2}) can be further simplified as 
\begin{equation}
\label{eq: APX_Int2-1}
 \lim_{a \rightarrow 0} \left[-\gamma(a) e^{-i\theta_a} \ln \left(\frac{b}{a}\right)\right] = \lim_{a \rightarrow 0} \left[\gamma(a) e^{-i\theta_a} \ln \left(a\right)\right] + \mathrm{O}\left(\ln \left(b\right)\right).
\end{equation}

Next, we move on to demonstrate that the second and the third terms of equation~(\ref{eq: APX_Int1}) are bounded. For the second term of equation~(\ref{eq: APX_Int1}), we again apply $z'-Z(a) = e^{i\theta_a} (s'-a)$ for $z' \in L_{ab}$ to obtain the relation,
\begin{equation}
\label{eq: APX_Int3}
 \frac{\gamma(s')-\gamma(a)}{Z_0-Z'(s')} = \frac{\gamma(s')-\gamma(a)}{Z_0 - Z(a) - e^{i\theta_a} (s'-a)}.
\end{equation}
Since $\gamma(s)$ is assumed to be infinitely differentiable for $a \leq s \leq b$, it can be expanded to Taylor series at $s=a$:
\begin{equation}
\label{eq: APX_Int4}
 \gamma(s) = \sum_{n=0}^{\infty} g_n(s-a)^{n} \hspace{5mm} \text{where} \hspace{5mm} g_n = \frac{\gamma^{(n)}(a)}{n!}.
\end{equation}
In the limiting case $a \rightarrow 0$, equation~(\ref{eq: APX_Int3}) can be combined with equation~(\ref{eq: APX_Int4}) to give
\begin{equation}
\label{eq: APX_Int5}
 \lim_{a \rightarrow 0} \frac{\gamma(s')-\gamma(a)}{Z_0-Z'(s')}\\
 = 
 \begin{dcases} 
   \lim_{a \rightarrow 0} \frac{0}{Z_0-Z(a)} = 0 \hspace{10mm} & \text{for} \hspace{5mm} s' = a,\\
   -e^{-i\theta_a} \sum_{n=0}^{\infty} g_{n+1} (s'-a)^{n} & \text{for} \hspace{5mm} a < s' \leq b.	 
 \end{dcases} 
\end{equation}
Since the series $\sum_{n=0}^{\infty} g_{n+1} (s'-a)^{n}$ converges uniformly to $(\gamma(s')-\gamma(a))e^{i\theta_a}/(Z'(s')-Z(a))$ for $a < s' \leq b$, its limit as $s' \rightarrow a$ takes the value
\begin{equation}
\label{eq: APX_Int6}
 \lim_{s' \rightarrow a} \sum_{n=0}^{\infty} g_{n+1} (s'-a)^{n} = \sum_{n=0}^{\infty} \lim_{s' \rightarrow a} \left[ g_{n+1} (s'-a)^{n} \right] = g_1,
\end{equation}
This immediately suggests that $\sum_{n=0}^{\infty} g_{n+1} (s'-a)^{n}$ is bounded when $s' \leq b$. Combining equations~(\ref{eq: APX_Int5}) and~(\ref{eq: APX_Int6}), we obtain
\begin{equation}
\label{eq: APX_Int7}
 \left| \lim_{a \rightarrow 0} \frac{\gamma(s')-\gamma(a)}{Z_0-Z'(s')} \right| < M_0 \hspace{5mm} \text{for} \hspace{5mm} a \leq s' \leq b,
\end{equation}
for a positive real number $M_0$. Therefore, the boundedness of the second term in equation~(\ref{eq: APX_Int1}) can be demonstrated by 
\begin{equation}
\label{eq: APX_Int8}
\begin{split}
 \lim_{a \rightarrow 0} \left|\int_{L_{ab}} \frac{\gamma(s')-\gamma(a)}{Z_0-Z'(s')} \mathrm{d}s' \right| &\leq \lim_{a \rightarrow 0} \int_{L_{ab}} \left| \frac{\gamma(s')-\gamma(a)}{Z_0-Z'(s')} \right| \mathrm{d}s'\\
 & < \lim_{a \rightarrow 0} \int_{L_{ab}} M_0 \mathrm{d}s' = M_0 b.
\end{split}   
\end{equation} 

Finally, we evaluate the boundedness of the third term in equation~(\ref{eq: APX_Int1}). As $X(s)$ and $Y(s)$ are infinitely differentiable for $a \leq s \leq b$, their Taylor series take the form 
\begin{align}
\label{eq: APX_Int9}
 X(s) = \sum_{n=0}^{\infty} f_{xn}(s-a)^{n} \hspace{5mm} \text{where} \hspace{5mm} f_{xn} = \frac{X^{(n)}(a)}{n!},\\
\label{eq: APX_Int10}
 Y(s) = \sum_{n=0}^{\infty} f_{yn}(s-a)^{n} \hspace{5mm} \text{where} \hspace{5mm} f_{yn} = \frac{Y^{(n)}(a)}{n!}.
\end{align}  
Denote $f_{zn} = f_{xn} + if_{yn}$, so equations~(\ref{eq: APX_Int9}) and~(\ref{eq: APX_Int10}) can be combined to give $Z(s) = \sum_{n=0}^{\infty} f_{zn}(s-a)^{n}$. Recall that $(\mathrm{d}Z'(s')/\mathrm{d}s')|_{s'=a} = (\mathrm{d}Z(s)/\mathrm{d}s)|_{s=a}$, it yields $Z'(s') = Z(a) + f_{z1}(s'-a)$ and the relation, 
\begin{equation}
\begin{split}
\label{eq: APX_Int11}
 \frac{1}{Z_0-Z(s')}-\frac{1}{Z_0-Z'(s')} &= \frac{Z(s')-Z'(s')}{(Z_0-Z(s'))(Z_0-Z'(s'))}\\
&= \frac{\sum_{n=2}^{\infty} f_{zn}(s'-a)^{n}}{[Z_0-Z(a)-\sum_{n=1}^{\infty} f_{zn}(s'-a)^{n}][Z_0-Z(a)-f_{z1}(s'-a)]},
\end{split}
\end{equation}  
which is associated with the third term of equation~(\ref{eq: APX_Int1}). As $a \rightarrow 0$, equation~(\ref{eq: APX_Int11}) has the limit
\begin{equation}
\label{eq: APX_Int12}
 \lim_{a \rightarrow 0} \left(\frac{1}{Z_0-Z(s')}-\frac{1}{Z_0-Z'(s')}\right)\\
 = 
 \begin{dcases} 
 \lim_{a \rightarrow 0} \frac{0}{(Z_0-Z(a))^2} = 0 \hspace{8mm} & \text{for} \hspace{5mm} s' = a,\\
 \frac{\sum_{n=2}^{\infty} f_{zn}(s'-a)^{n}}{\sum_{n=1}^{\infty} f_{zn}f_{z1}(s'-a)^{n+1}} & \text{for} \hspace{5mm} a < s' \leq b. 
\end{dcases}
\end{equation}
Similar to the derivation of the second term of equation~(\ref{eq: APX_Int1}), the series $\sum_{n=2}^{\infty} f_{zn}(s'-a)^{n}$ and $\sum_{n=1}^{\infty} f_{zn}f_{z1}(s'-a)^{n+1}$ converge uniformly to $Z(s')-Z'(s')$ and $(Z(a)-Z(s'))(Z(a)-Z'(s'))$, respectively, for $a < s' \leq b$. As a result, the limit of $\sum_{n=2}^{\infty} f_{zn}(s'-a)^{n}/\sum_{n=1}^{\infty} f_{zn}f_{z1}(s'-a)^{n+1}$ approaches $f_{z2}/f_{z1}^2$ as $s' \rightarrow a$. Therefore, equations~(\ref{eq: APX_Int11}) and~(\ref{eq: APX_Int12}) together indicate that
\begin{equation}
\label{eq: APX_Int13}
 \left| \lim_{a \rightarrow 0} \left(\frac{1}{Z_0-Z(s')}-\frac{1}{Z_0-Z'(s')}\right) \right| < M_1 \hspace{5mm} \text{for} \hspace{5mm} a \leq s' \leq b,
\end{equation}
where $M_1$ is also a positive value. It is apparent that there exists a positive value $M_2$ so that $|\gamma(s')| < M_2$ for $a \leq s' \leq b$. As a result, the boundedness of third term in equation~(\ref{eq: APX_Int1}) can be reflected from the following inequality, 
\begin{equation}
\label{eq: APX_Int14}
\begin{split}
 &\lim_{a \rightarrow 0} \left|\int_{L_{ab}} \left( \frac{1}{Z_0-Z(s')}-\frac{1}{Z_0-Z'(s')} \right) \gamma(s') \mathrm{d}s' \right|\\
 & \leq \lim_{a \rightarrow 0} \int_{L_{ab}} \left| \frac{1}{Z_0-Z(s')}-\frac{1}{Z_0-Z'(s')} \right| \left|\gamma(s')\right| \mathrm{d}s'\\
 & < \lim_{a \rightarrow 0} \int_{L_{ab}} M_1M_2 \mathrm{d}s' = M_1M_2 b.
\end{split}   
\end{equation}
Combing equations~(\ref{eq: APX_Int1}),~(\ref{eq: APX_Int2-1}),~(\ref{eq: APX_Int8}), and~(\ref{eq: APX_Int14}), we obtain the result of equations~(\ref{eq: APX_Int0}) in the form, $\gamma(a) e^{-i\theta_a} \ln(a) + \mathrm{O}(\ln(b)) + \mathrm{O}(b)$ as $a \rightarrow 0$. 
		
\section{} \label{sec:11}

This appendix is a continuation of appendix~\ref{sec:9}. Appendix~\ref{sec:9} has calculated the integral $\int \gamma(s)(Z_0-Z(s))^{-1} \mathrm{d}s$ between $Z(a)$ and $Z(b)$ for the limiting case $Z(a)\rightarrow Z_0$. In this part, the goal is to prove that the original integral is bounded if $a$ is a finite value or $Z(a)$ is away from $Z_0$. 

For this purpose, we only require $\gamma(s)$ to be bounded for $a \leq s \leq b$. Let $D_m = \min\{|Z-Z_0|, Z \in C_{ab}\}$ and $\gamma_M = \max\{\gamma(s), s \in [a,b]\}$, the original integral has the inequality
 \begin{equation}
\label{eq: APX1_Int1}
\begin{split}
 &\left|\int_{C_{ab}} \frac{\gamma(s) \mathrm{d}s}{Z_0-Z(s)}  \right| \leq \int_{C_{ab}} \left| \frac{1}{Z_0-Z(s)} \right| \left|\gamma(s)\right| \mathrm{d}s < \int_{C_{ab}} \frac{\gamma_M}{D_m} \mathrm{d}s = \frac{\gamma_M(b-a)}{D_m}.
\end{split} 
\end{equation} 
Clearly, this proves the boundedness of the original integral.    

 \newcommand{\AIAAJ}{AIAA J.} \newcommand{\AIAAP}{AIAA Paper}
  \newcommand{\ARMA}{Archive for Rational Mechanics and Analysis}
  \newcommand{\ASMEJFE}{J. Fluids Eng., Trans. ASME} \newcommand{\ASR}{Applied
  Scientific Research} \newcommand{\CF}{Computers Fluids}
  \newcommand{\CJFAS}{Can. J. Fish. Aquat. Sci.}
  \newcommand{\ETFS}{Experimental Thermal and Fluid Science}
  \newcommand{\EF}{Experiments in Fluids} \newcommand{\FDR}{Fluid Dynamics
  Research} \newcommand{\IJHMT}{Int. J. Heat Mass Transfer}
  \newcommand{\JASA}{J. Acoust. Soc. Am.} \newcommand{\JCP}{J. Comp. Physics}
  \newcommand{\JEB}{J. Exp. Biol.} \newcommand{\JFM}{J. Fluid Mech.}
  \newcommand{\JMP}{J. Math. Phys.} \newcommand{\JSC}{J. Scientific Computing}
  \newcommand{\JSP}{J. Stat. Phys.} \newcommand{\JSV}{J. of Sound and
  Vibration} \newcommand{\MC}{Mathematics of Computation}
  \newcommand{\MWR}{Monthly Weather Review} \newcommand{\PAS}{Prog. in
  Aerospace. Sci.} \newcommand{\PCPS}{Proc. Camb. Phil. Soc.}
  \newcommand{\PD}{Physica D} \newcommand{\PRA}{Physical Rev. A}
  \newcommand{\PRE}{Physical Rev. E} \newcommand{\PRL}{Phys. Rev. Lett.}
  \newcommand{\PF}{Phys. Fluids} \newcommand{\PFA}{Phys. Fluids A.}
  \newcommand{\PL}{Phys. Lett.} \newcommand{\PRSLA}{Proc. R. Soc. Lond. A}
  \newcommand{\SIAMJMA}{SIAM J. Math. Anal.} \newcommand{\SIAMJNA}{SIAM J.
  Numer. Anal.} \newcommand{\SIAMJSC}{SIAM J. Sci. Comput.}
  \newcommand{\SIAMJSSC}{SIAM J. Sci. Stat. Comput.}
  \newcommand{\TCFD}{Theoret. Comput. Fluid Dynamics} \newcommand{\ZAMM}{ZAMM}
  \newcommand{\ZAMP}{ZAMP} \newcommand{\ICASER}{ICASE Rep. No.}
  \newcommand{\NASACR}{NASA CR} \newcommand{\NASATM}{NASA TM}
  \newcommand{\NASATP}{NASA TP} \newcommand{\ARFM}{Ann. Rev. Fluid Mech.}
  \newcommand{\WWW}{from {\tt www}.} \newcommand{\CTR}{Center for Turbulence
  Research, Annual Research Briefs} \newcommand{\vonKarman}{von Karman
  Institute for Fluid Dynamics Lecture Series}

\end{document}